\documentclass[ammsymb, twocolumn, prb, superscriptaddress, floats, nobibnotes, a4paper]{revtex4-1}

\usepackage[T1]{fontenc}
\usepackage{bm}
\usepackage{graphicx}
\usepackage{natbib}
\usepackage{epstopdf}
\usepackage[euler]{textgreek}
\usepackage{hyperref}
\usepackage{color}

\usepackage[caption=false]{subfig}

\setcitestyle{numbers,square}

\begin{document}

\title{Brightening of dark excitons in monolayers of semiconducting transition metal dichalcogenides}

\author{M. R. Molas}
\email{maciej.molas@gmail.com}
\affiliation{Laboratoire National des Champs Magn\'etiques Intenses, CNRS-UGA-UPS-INSA-EMFL, 25, avenue des Martyrs, 38042 Grenoble, France}
\author{C. Faugeras}
\affiliation{Laboratoire National des Champs Magn\'etiques Intenses, CNRS-UGA-UPS-INSA-EMFL, 25, avenue des Martyrs, 38042 Grenoble, France}
\author{A. O. Slobodeniuk}
\affiliation{Laboratoire National des Champs Magn\'etiques Intenses, CNRS-UGA-UPS-INSA-EMFL, 25, avenue des Martyrs, 38042 Grenoble, France}
\author{K. Nogajewski}
\affiliation{Laboratoire National des Champs Magn\'etiques Intenses, CNRS-UGA-UPS-INSA-EMFL, 25, avenue des Martyrs, 38042 Grenoble, France}
\author{M. Bartos}
\affiliation{Laboratoire National des Champs Magn\'etiques Intenses, CNRS-UGA-UPS-INSA-EMFL, 25, avenue des Martyrs, 38042 Grenoble, France}
\author{D. M. Basko}
\affiliation{Laboratoire de Physique et Mod\'elisation des Milieux Condens\'es, Universit\'e de Grenoble-Alpes and CNRS, 25 rue des Martyrs, 38042 Grenoble, France}
\author{M. Potemski}
\email{marek.potemski@lncmi.cnrs.fr}
\affiliation{Laboratoire National des Champs Magn\'etiques Intenses, CNRS-UGA-UPS-INSA-EMFL, 25, avenue des Martyrs, 38042 Grenoble, France}

\date{\today}

\begin{abstract}

We present low temperature magneto-photoluminescence experiments
which demonstrate the brightening of dark excitons by an in-plane
magnetic field $B$ applied to monolayers of different semiconducting
transition metal dichalcogenides. For both WSe$_2$ and  WS$_2$
monolayers, the dark exciton emission is observed at $\sim$50~meV
below the bright exciton peak and displays a characteristic
doublet structure which intensity is growing with $B^2$, while no
magnetic field induced emission peaks appear for MoSe$_2$
monolayer. Our experiments also show that the MoS$_2$ monolayer
has a dark exciton ground state with a dark-bright exciton
splitting energy of $\sim$100~meV.

\end{abstract}

\maketitle

\section{Introduction}

Monolayers (MLs) of semiconducting transition metal
dichalcogenides (S-TMDs) MX$_2$ where M$=$Mo or W and X$=$S, Se or
Te, are direct band gap semiconductors~\cite{mak2010} with the
minima (maxima) of conduction (valence) band located at the
inequivalent K$^+$ and K$^-$ points of their hexagonal Brillouin
zone (BZ). These two-dimensional semiconductors host tightly bound
excitons with unconventional properties such as binding energies
as large as few hundreds of meV and non Rydberg excitation
spectrum~\cite{ugeda2014, ye, chernikov2015}. The lack of inversion
symmetry together with the strong spin-orbit interaction lift the
degeneracy between spin levels in the conduction (CB) and valence
(VB) bands at the K$^+$ and K$^-$ points related by time reversal
symmetry. The spin-orbit interaction leads to well separated spin
subbands in each valley and to the possibility of initializing a
defined valley population with circularly polarized optical
excitation~\cite{xiao,mak2012,cao,zeng2012} or generation of
valley coherence~\cite{jones,wang2016}. The spin-orbit splitting
$\Delta_{so,vb}$ in the valence is as large as few hundreds of
meV~\cite{britton,zhang,riley2014,ross,ye,chernikov,zhao2012,kozawa,zeng,liPRB,zhu,klots,hanbicki}
while its counterpart in the conduction band $\Delta_{so,cb}$ is
predicted to be of the order of few tens of meV only. What is
however important is that $\Delta_{so,cb}$ can be positive or
negative~\cite{Kosmider2013a, Kosmider2013b} and in consequence,
two distinct ordering of the spin orbit split CB subbands are
feasible~\cite{liu,kormanyos,Echeverry2016}.

\begin{center}
    \begin{figure*}[t]
        \centering
        \includegraphics[width=0.7\linewidth]{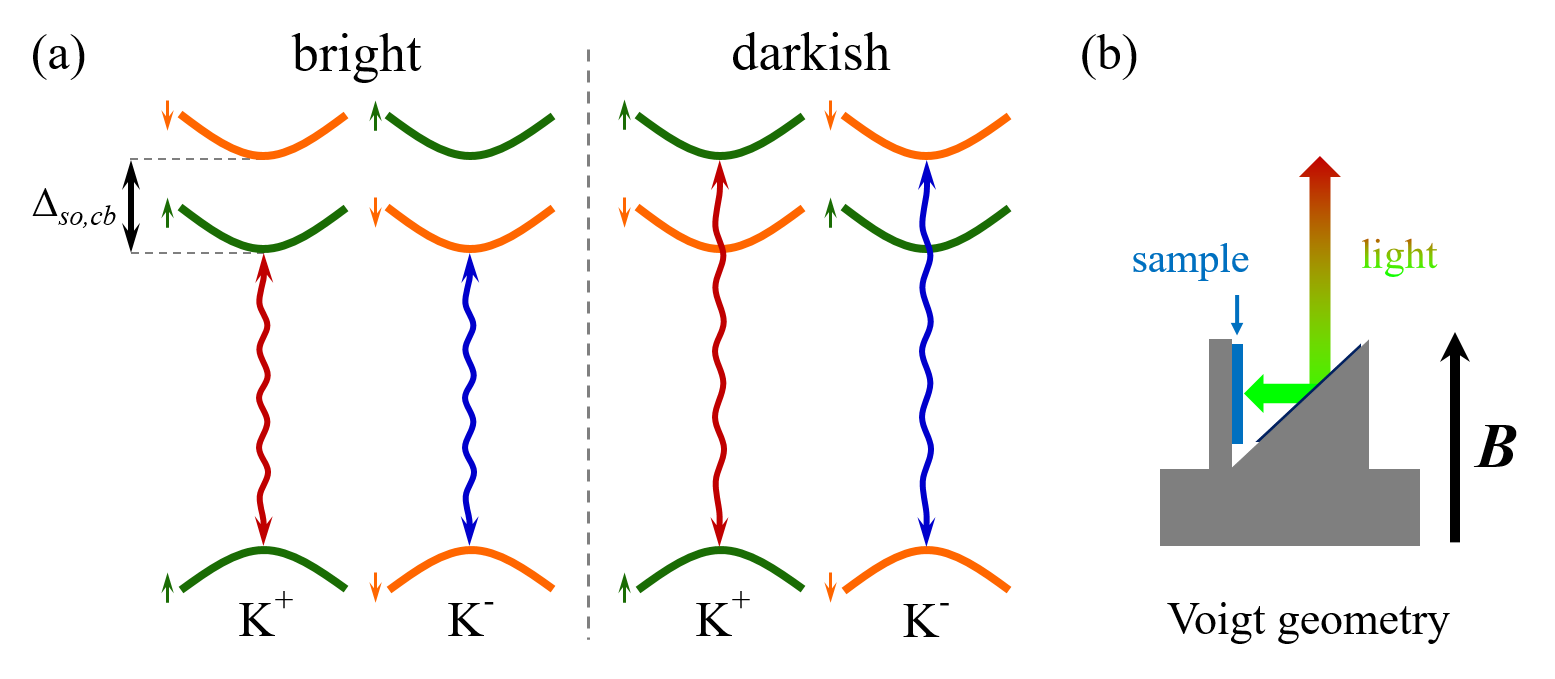}%
        \caption{(a) Diagram of relevant subbands in the CB and VB at the K$^+$ and K$^-$ points of the BZ in the bright and darkish monolayers of S-TMDs.
        The orange (green) curves indicate the spin-up (spin-down) subbands. The red and blue wavy lines show the A exciton transitions which are optically active.
        $\Delta_{so,cb}$ denotes the spin-orbit splitting in the conduction band.
        (b) Schematic representation of the experimental configuration for magneto-PL measurements in Voigt configuration.}
        \label{fig:theory}
    \end{figure*}
\end{center}

Because optical transitions in S-TMDs do conserve the spin,
different orderings of electronic bands in the conduction band
have profound consequences on their optical properties. Depending
on the sign of $\Delta_{so,cb}$, the excitonic ground state can be
bright (parallel spin configuration for the top VB and the lowest
CB subbands between which the optical transition is allowed) or
dark (anti-parallel spin configuration and optically forbidden
ground state interband transition). The ordering of the electronic
bands, characteristic for these two monolayer families, referred
to as bright and darkish ones, are illustrated in
Fig.~\ref{fig:theory}(a). Theoretical studies~\cite{liu,kormanyos,Echeverry2016} indeed predict that monolayers
of MoSe$_2$ and of MoTe$_2$ should be bright ($\Delta_{so,cb}$>0
to set a convention) while WSe$_2$ and WS$_2$ monolayers are
darkish ($\Delta_{so,cb}$<0). Yet, there is no general consensus
concerning the bright of darkish character of a MoS$_2$ monolayer.
The theoretical works reported in
Refs~\citenum{kormanyos,Echeverry2016} classify the MoS$_2$ monolayer
as a bright system: $\Delta_{so,cb}$>0 but as small as
$3$~meV. Instead, another
theoretical study~\cite{Qiu2015} indicates that MoS$_2$
monolayers are rather darkish ($\Delta_{so,cb}\sim -40meV$).

A detailed knowledge of the exciton fine structure is crucial for
S-TMD based optoelectronic devices and for valleytronic
applications, as i) optical properties strongly depend on the type
of excitonic ground state, and ii) scattering mechanisms, and in
particular intervalley scattering mechanisms, can have much
different efficiencies for bright and dark
excitons~\cite{Glazov2014,smolenski}. On the experimental point of
view, recent optical studies of WSe$_2$ have shown that the
temperature dependence of its PL intensity is consistent with a
dark excitonic ground state and the dark-bright exciton splitting
of about $30$~meV has been experimentally estimated from
temperature activation type
analysis~\cite{arora,Zhang2015a,Wang2015a}.

Magnetic fields, applied in an adequate configuration with respect
to a crystal axis, can mix electronic wave functions and thus the
excitonic states which are built out of these wave functions. This
effect triggered the spectroscopy of optically dark excitons in a
large variety of condensed matter systems, ranging from bulk
semiconductors~\cite{Brandt2007}, semiconductor quantum
dots~\cite{Nirmal1995,Bayer2000}, to single wall carbon
nanotubes~\cite{Zaric2004,Srivastava2008}. One expects that also
in monolayers of S-TMDs, the in-plane magnetic field acts as a
perturbation to the system's Hamiltonian, mixing the two lowest
spin levels in the CB and hence, the bright excitons giving some
optical activity to the initially dark excitonic
states~\cite{Slobo2016}.

In this paper, we provide a direct measurement of the dark exciton
emission in darkish monolayers of S-TMDs by mixing the spin levels
of bright and dark excitons by an in-plane magnetic field. Dark
excitons appear in the low temperature magneto-photoluminescence
(PL) spectra as clear features growing with the magnetic field at
energies lower than that of the bright exciton. This observation
gives a direct access to the values of dark-bright exciton
splitting in WSe$_2$ and WS$_2$ monolayers. In the case of
MoSe$_2$, no significant change in the emission spectrum is
observed when applying a magnetic field, in agreement with its
bright exciton ground state. MoS$_2$ is shown to belong to the
family of darkish materials with a dark-bright splitting energy
close to $100$~meV. The emission intensity of dark excitons
increases as $B^2$, in accordance with their perturbative
activation by the in-plane magnetic field.

\section{Theoretical Background}

To examine the band edge interband transitions in S-TMD monolayers
we consider the top VB and two spin-orbit split CB subbands [see
Fig.~\ref{fig:theory}(a)]. Associated with these subbands and
relevant for our considerations are intravalley interband
transitions (intravalley A excitons) which involve the states from
the the same K$^+$ or K$^-$ valley. Four types of intravalley A
excitons can be distinguished and labelled according to their
valley $\tau=\pm$ and CB spin $s_{cb}=\uparrow,\downarrow$ indices
(the VB spin index, $s_{vb}$ is fixed to $s_{vb}=\uparrow$ in
$\mathbf{K}^+$ valley and $s_{vb}=\downarrow$ in $\mathbf{K}^-$
valley). The configurations with $s_{vb}=s_{cb}=\uparrow$ from the
K$^+$ valley and $s_{vb}=s_{cb}=\downarrow$ from the K$^-$ valley
correspond to \textit{optically active}, bright A excitons,
referred to as $|\tau, \mathrm{b}\rangle$. The configurations with
$s_{vb}=\uparrow, s_{cb}=\downarrow$ from the K$^+$ valley and
$s_{vb}=\downarrow, s_{cb}=\uparrow$ from the K$^-$ valley
correspond to \textit{optically inactive}, dark A excitons,
refereed to as $|\tau, \mathrm{d}\rangle$.

The ground state of bright (darkish) S-TMD monolayers is then
formed from bright (dark) A excitons. Because these
quasi-particles differ from each other by their spin configuration
in the CB, spin-flip processes in the CB can make the dark states
optically active and can allow for investigations of the ground
state of darkish materials.

A magnetic field $\mathbf{B}=(B_x,B_y)$, applied along the plane
of a S-TMD monolayer, mixes the spin states in the CB and VB via
the Zeeman interaction. Since $\Delta_{so,vb} \gg \Delta_{so,cb}$,
the spin-mixing in the VB can be neglected. The Zeeman term acting
on the CB states can be expressed as:

\begin{equation}
H_Z=\frac12g_{cb}\mu_\mathrm{B}(\sigma_x B_x +\sigma_y B_y)
\end{equation}

Here $g_{cb}$ is the in-plane gyromagnetic ratio for the CB,
$\mu_\mathrm{B}$ is the Bohr magneton and $\sigma_{x,y}$ are the
Pauli matrices in the CB spin subspace. The in-plane magnetic
field results in the mixing of the dark and bright excitons. It
can be described by an effective $2\times{2}$ Hamiltonian in the
basis of \{$|\tau,\mathrm{b}\rangle$, $|\tau,\mathrm{d}\rangle$\},
obtained by the projection of spin states of the CB, mixed by the
magnetic field, on the exciton states

\begin{equation}
H^\tau_\mathrm{ex}=
\left[\begin{array}{cc}
E_\mathrm{b} &  \frac12g_{cb}\mu_\mathrm{B}B_{-\tau}  \\
\frac12g_{cb}\mu_\mathrm{B}B_\tau & E_\mathrm{d}
\end{array}\right].
\end{equation}

Here we introduced $B_\pm=B_x\pm iB_y$. $E_\mathrm{b}$ and
$E_\mathrm{d}$ are the energies of the bright and dark excitons in
the absence of an external magnetic field,
$E_\mathrm{d}-E_\mathrm{b} = \Delta_{so,cb}$. The application of
the in-plane magnetic field does not lift the double degeneracy of
each dark and bright exciton states as $H^\tau_\mathrm{ex}$ does
not depend on valley index $\tau$.

Assuming that the Zeeman term gives a small correction to the
basic exciton states, we obtain the mixed eigenstates up to second
order in magnetic field:

\begin{eqnarray}
|\tau,\mathrm{b}\rangle_\mathrm{mix}\!&=&\!
\frac{|\tau,\mathrm{b}\rangle}{1+w/2} -
\frac{g_{cb}\mu_\mathrm{B}B_\tau}{2\Delta_{so,cb}}
|\tau,\mathrm{d}\rangle,\\
|\tau,\mathrm{d}\rangle_\mathrm{mix}\!&=&\!
\frac{|\tau,\mathrm{d}\rangle}{1+w/2}
+\frac{g_{cb}\mu_\mathrm{B}B_{-\tau}}{2\Delta_{so,cb}}
|\tau,\mathrm{b}\rangle.
\end{eqnarray}

Here $w=g_{cb}^2\mu_\mathrm{B}^2 B^2/(4\Delta_{so,cb}^2)\ll 1$.
Their eigenenergies are very close to the energies of the dark and
bright excitons (the correction is $\propto w\Delta_{so,cb}$).

The admixture of bright states to the dark exciton state makes the
latter resonance to be possibly observed in the PL spectra when
the in-plane magnetic field is applied to the layer. The intensity
$I_\mathrm{d}$ of such a PL line can be expected to be
proportional to the fraction $w$ of bright exciton in the
corresponding mixed state and to the population $n_\mathrm{d}$ of
dark excitons:

\begin{equation}
I_\mathrm{d}=n_\mathrm{d}I_\mathrm{b}w \propto
n_\mathrm{d}I_\mathrm{b} B^2, \label{Zeequation}
\end{equation}

where $I_\mathrm{b}$ is the intensity of the pure bright exciton
state emission in the absence of the magnetic field. With
available magnetic fields, the factor $w$ remains rather small and
dark excitons can hardly be observed in absorption experiments. We
note two different situations. i)  For bright materials, such as
MoSe$_2$ or MoTe$_2$, the energy of dark excitons is larger than
the energy of the bright ones. Therefore, at low temperatures, the
population of dark excitons is suppressed by a Boltzmann factor
$exp(-\Delta_{so,cb}/k_BT)$ and optical transitions are mainly due
to low-lying bright exciton states. In this case the observation
of dark excitons at low temperature is extremely unlikely. ii) For
darkish materials, such as WSe$_2$ or WS$_2$, the situation is
opposite and the direct observation of dark excitons is possible.

So far we have considered the A excitons formed by the direct
electron-hole Coulomb interaction and have not included effects of
the exchange part of the Coulomb interaction. The exchange
interaction is expected to lift the double valley degeneracy of
dark intravalley A excitons due to the presence of a transition
dipole moment perpendicular to the monolayer plane, absent for
bright excitons~\cite{dery2015,Slobo2016}. This degeneracy lifting
can be viewed as a local-field effect due to the out-of-plane
transition dipole moment of spin-forbidden dark excitons. It is
analogous to the exchange energy shift of the Z-excitons in
semiconductor quantum wells~\cite{chen1988,andreani1990}. The
resulting energy splitting between the two spin-forbidden dark
exciton components in S-TMD monolayers was roughly estimated in
Ref.~\citenum{Slobo2016} to be about 10 meV; a more precise,
microscopic calculation of this splitting is still lacking, to the
best of our knowledge. The discussed above effects of the in-plane
magnetic field are equally valid for each component of the
expected doublet structure of dark excitons in S-TMD monolayers.

\begin{center}
    \begin{figure*}[!t]
        \subfloat{}%
        \centering
        \includegraphics[width=0.5\linewidth]{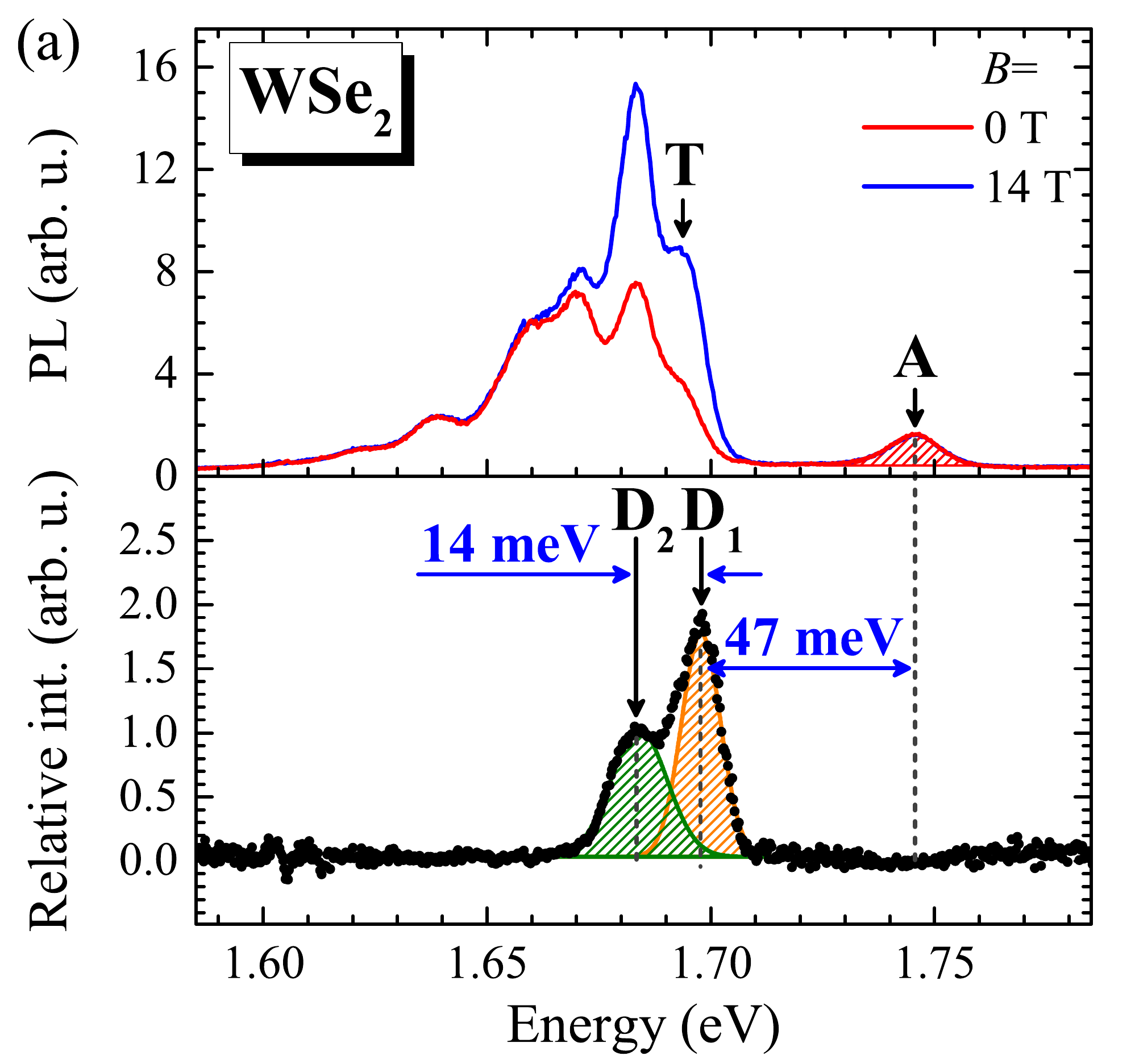}%
        \subfloat{}%
        \centering
        \includegraphics[width=0.5\linewidth]{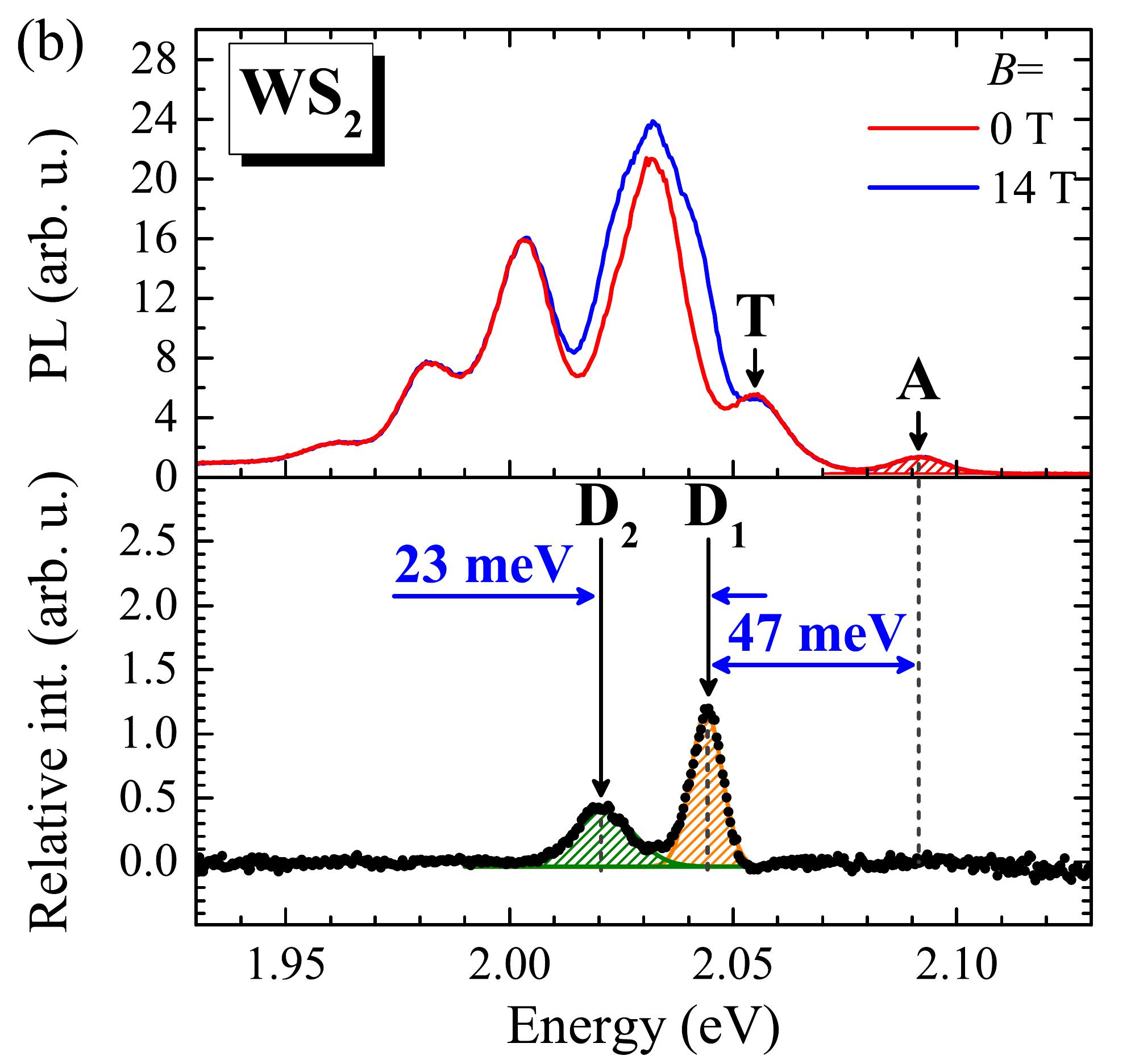}%

        \subfloat{}%
        \centering
        \includegraphics[width=0.5\linewidth]{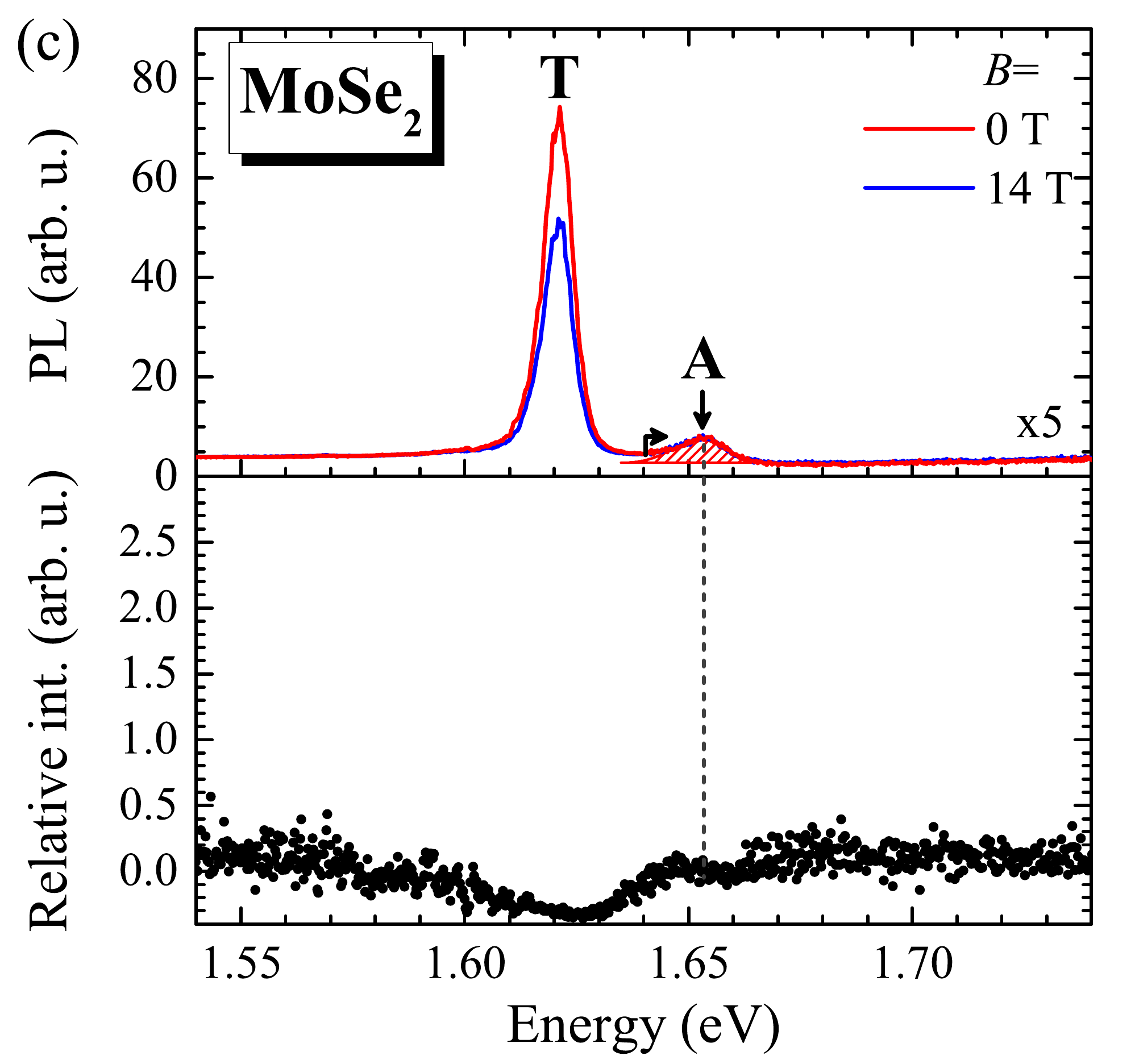}%
        \subfloat{}%
        \centering
        \includegraphics[width=0.5\linewidth]{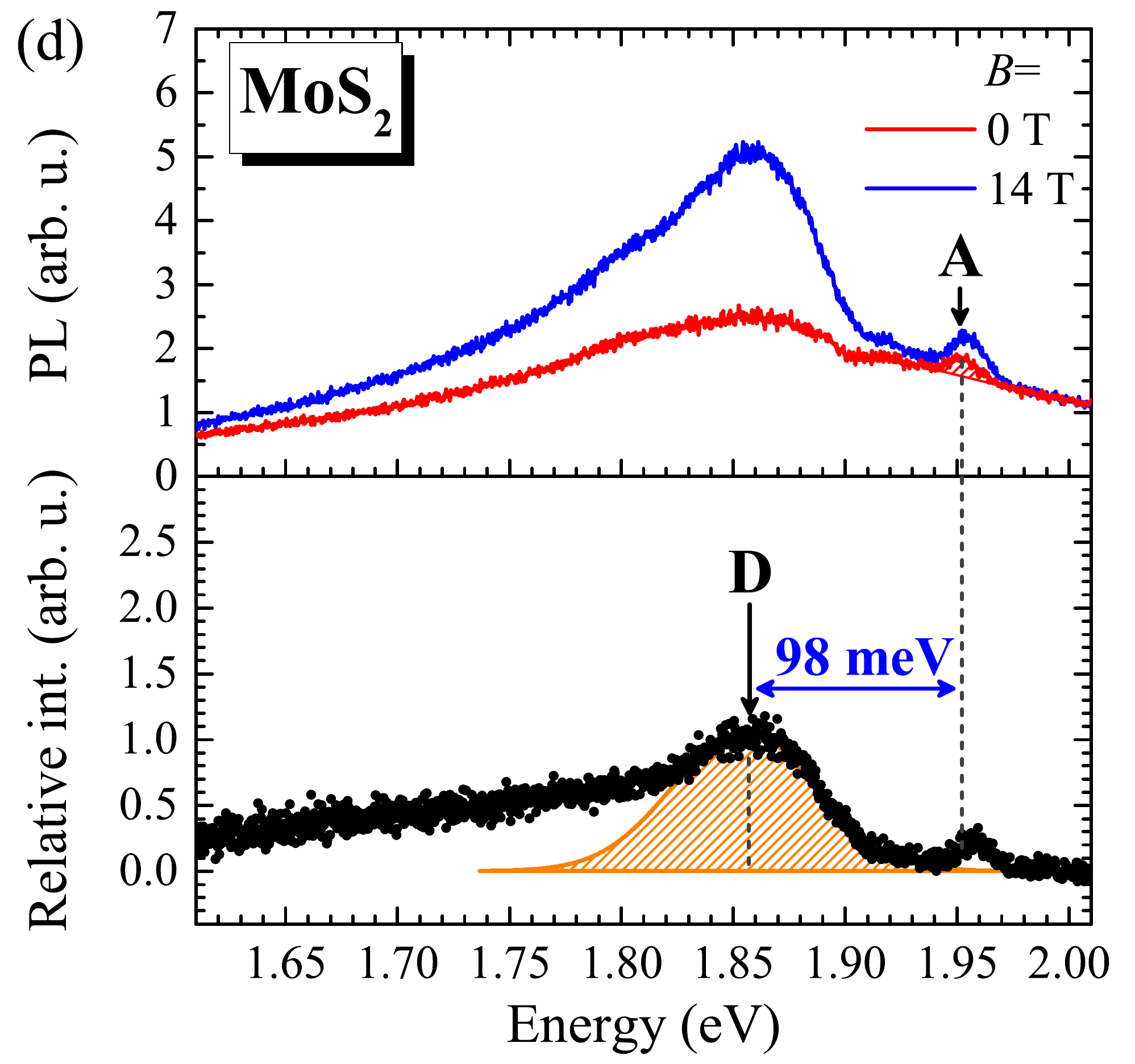}%

        \caption{(upper panels) PL spectra of (a) WSe$_2$, (b) WS$_2$, (c) MoSe$_2$, and (d) MoS$_2$ monolayers at $T=4.2$~K measured at zero field (red curves) and at $B=14$~T (blue curves) applied in the plane of the crystal. The PL spectra were normalized to the intensity of the A exciton line. (lower panels) Corresponding relative intensities of the monolayers defined as (PL$_{B=\textrm{14 T}}$ - PL$_{B=\textrm{0 T}}$)/PL$_{B=\textrm{0 T}}$ are represented by black dots. The orange and green curves indicate Gaussians fits of the data.}
        \label{fig:pl}
    \end{figure*}
\end{center}

\section{Experimental results and Discussion}

Monolayers of S-TMDs have been prepared by mechanical exfoliation
of bulk crystals purchased from HQ Graphene. Initially, the flakes
were exfoliated onto a polydimethylsiloxane (PDMS) stamp attached
to a glass plate. MLs of S-TMDs were then identified by their
optical contrast and cross-checked by Raman scattering and PL
measurements at room temperature. In order to deposit them on
target Si/SiO$_2$(320 nm) substrates, an all-dry PDMS-based
transfer method similar to the one described in Ref.~\citenum{gomez}
was employed.

Low temperature magneto-PL experiments were performed in the Voigt
configuration [see Fig.~\ref{fig:theory}(b)] using an
optical-fiber-based insert placed in a superconducting magnet
producing magnetic fields up to $14$~T. The samples were placed on
top of a x-y-z piezo-stage kept in gaseous helium at $T=4.2$~K.
The light from a semiconductor diode laser ($\lambda$=515~nm) was
coupled to an optical fiber with a core of 50 $\mu$m diameter and
focused on the sample by an aspheric lens (spot diameter around 10~$\mu$m). PL signals were collected by the same lens, injected into
a second optical fiber of the same diameter, and analyzed by a
$0.5$~m long monochromator equipped with a charge-couple-device
(CCD) camera.

To investigate the effect of an in-plane magnetic field on the PL
signal of S-TMD monolayers, we measured the evolution of the low
temperature ($T=4.2$~K) PL spectra of the WSe$_2$, WS$_2$,
MoSe$_2$, and MoS$_2$ MLs in the Voigt configuration as a function
of an external magnetic field up to $B=14$~T. The obtained spectra
at $B=0$ and at $B=14$~T are presented in the upper panels of
Fig.~\ref{fig:pl}. The zero-field PL spectra of all our monolayers
display two characteristic emission features, labelled A and T,
which are associated with recombination of the neutral [an
electron-hole ($eh$) pair] and charged [an $eh$ pair + an extra
carrier (electron or hole)] excitons formed at the K$^\pm$ points
of the
BZ~\cite{jones,arora,mitoglunano,smolenski,plechinger,shang,li2014,aroramose2,cadiz,cadizacid}.
In the case of WSe$_2$, WS$_2$, and also of MoS$_2$, additional
features are apparent in the PL spectra in the form of a series of
emission lines (WSe$_2$ and WS$_2$) or a broad band (MoS$_2$), at
energies below the A exciton energy and overlapping with the T
peak. These additional lines have been attributed in the
literature to the so-called localized/bound or defect-related
excitons~\cite{jones,arora,mitoglunano,smolenski,plechinger,shang}.

We start with the analysis of the results obtained for the
tungsten-based family, $i.e.$ WSe$_2$ and WS$_2$ MLs, as both of
them are rather firmly predicted to belong to the family of
darkish monolayers~\cite{liu,kormanyos,Echeverry2016}. The
zero-field PL spectra, apart the A and T peaks, consist of several
overlapping emission lines on the lower energy side of the
spectrum [upper panels of Fig.~\ref{fig:pl}(a) and (b)]. We show
in Fig.~\ref{fig:pl}(a) and (b) that the application of a magnetic
field in the plane of these monolayers strongly affects their PL
spectra at energies $50 - 60$~meV below the A exciton line. To
better visualize the effects of magnetic fields and compare the
results obtained for different materials, we define a relative
spectrum as (PL$_{B\neq\textrm{0}}$ -
PL$_{B=\textrm{0}}$)/PL$_{B=\textrm{0}}$. Such relative intensity
spectra for $B=14$~T are presented in the lower panels of
Fig.~\ref{fig:pl}(a) and (b). For WSe$_2$ and WS$_2$ MLs, these
spectra are composed of two peaks, labelled D$_1$ and D$_2$, which
appear on the lower energy side of the bright A exciton. In
agreement with our theoretical arguments, these two peaks are
assigned to the magnetic-field induced emission due to dark
excitons. The higher energy peak, D$_1$, emerges about $47$~meV
below the A exciton line for both members of the tungsten-based
family. The energy separation between the D$_1$ and D$_2$ peaks is
$14$~meV for WSe$_2$ monolayer and $23$~meV for the WS$_2$
monolayer.

\begin{center}
    \begin{figure}[]
        \centering
        \includegraphics[width=1\linewidth]{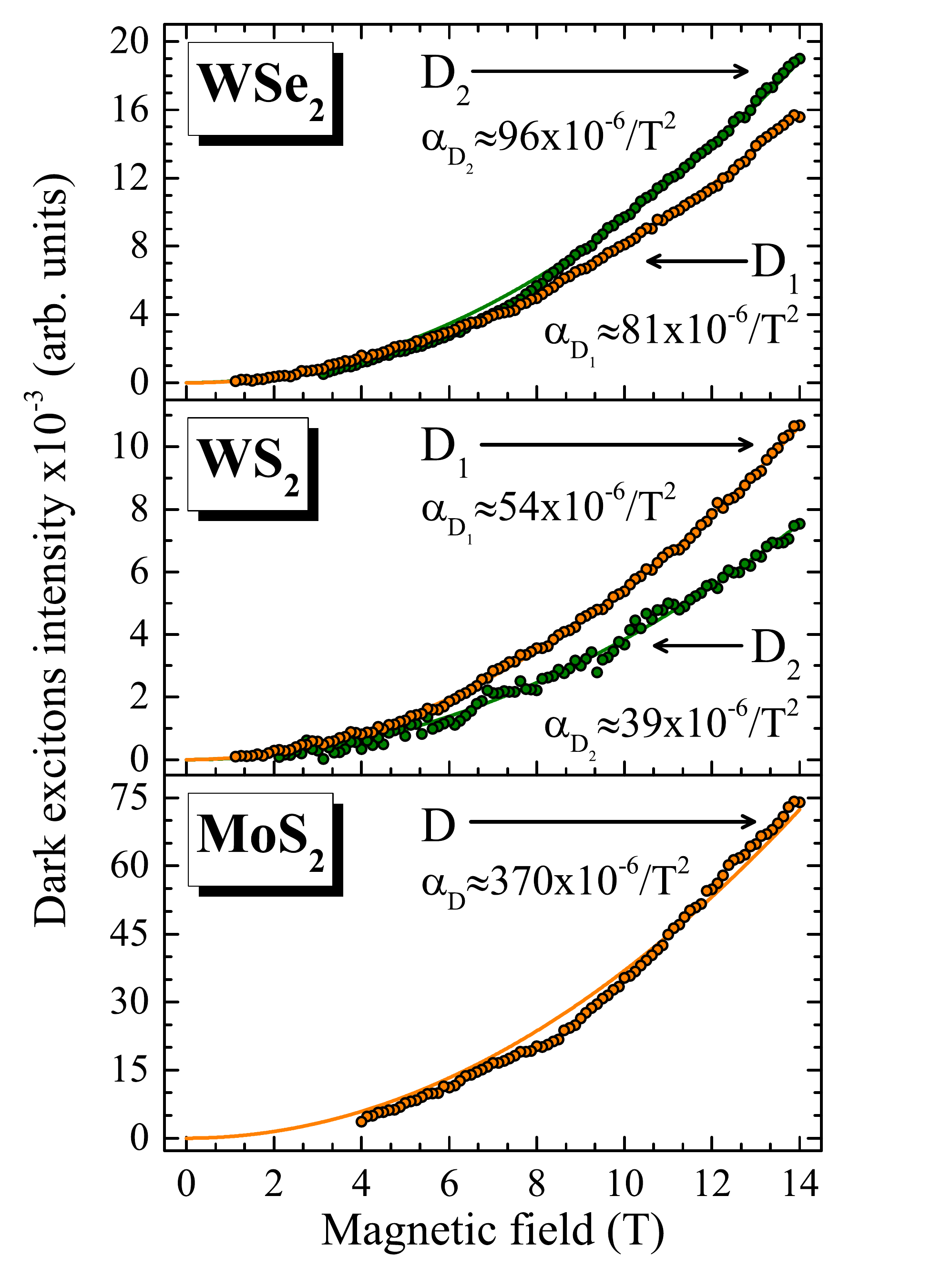}%
        \caption{Magnetic-field dependence of the intensities of dark exciton lines (D$_1$, D$_2$, and D) obtained on WSe$_2$, WS$_2$, and MoS$_2$ monolayers.}
        \label{fig:dark}
    \end{figure}
\end{center}

To analyze further the data, we fitted the D$_1$ and D$_2$
features using two Gaussian functions [see lower panels of
Fig.~\ref{fig:pl} (a) and (b)]. In the whole range of investigated
magnetic fields, the energy and the full width at half maximum
(FWHM) of the two D$_1$ and D$_2$ peaks are constant. The
brightening of these dark excitons is evidenced by the quadratic
evolution of the integrated intensity of these peaks as a function
of the magnetic field ($\sim\alpha B^2$, where $\alpha$ is a
fitting parameter). This behavior is presented in
Fig.~\ref{fig:dark} and is in agreement with the arguments
presented in the preceding section (Eq.~\ref{Zeequation}).
Important here is the observed $B^2$ dependence and not the
precise rates of increase of the two D lines, which apparent
values are affected by the chosen normalization of the relative
spectra. We consider that the energy difference between the bright
A exciton peak and the dark D$_1$ exciton peaks corresponds well
to the theoretical predictions of the $\Delta_{so,cb}$
magnitude~\cite{liu,kormanyos,Echeverry2016}. Note that the values
for $\Delta_{so,cb}$ calculated in
Ref.~\citenum{liu,kormanyos,Echeverry2016} do not include the
electron-hole Coulomb effects, which obviously  affect the
interband transition energies but can also significantly influence
the apparent bright-dark exciton splitting due to electron-hole
exchange effects.

An MoSe$_2$ monolayer is predicted to belong to the family of
bright S-TMDs. Its zero-field PL spectra is rather simple (and
similar to that observed for MoTe$_2$
monolayers~\cite{Ignacio2015}). It is composed of only two A and T
features [see Fig.~\ref{fig:pl}(c)]\cite{li2014,aroramose2,cadiz}.
When a magnetic field is applied in the direction along the plane
of the layer, no significant changes of the PL spectra are
observed. In particular, there are no additional growing
structures on the high energy side of the A exciton line, where
the dark exciton emission could be expected according to the band
ordering at the K$^\pm$ points [see Fig.~\ref{fig:theory}(a)]. The
dark exciton emission can not be detected with our experimental
conditions as a result of the fast relaxation of carriers to the
lowest energy state which is a bright exciton. The only field
induced effect observed in the magneto-PL spectra of the MoSe$_2$
monolayer is a small decrease in the intensity of the T-peak [see
lower panel of Fig.~\ref{fig:pl}(c)]. The origin of this field
induced suppression of the trion emission is not clear for us and
calls for a possible theoretical explanation, thought one may
speculate that it reflects an influence of the magnetic field on
the formation of the charged excitons in a MoSe$_2$ monolayer
through a mixing of the spin split bands in both valleys.

Existing models describing the band ordering for MoS$_2$ monolayer
largely predict a positive, thought small, ${\sim3}$~meV  value for
$\Delta_{so,cb}$~\cite{liu,kormanyos,Echeverry2016}, thus placing
MoS$_2$ in the family of bright materials. Recently, however, a
negative dark-bright exciton splitting has been
predicted~\cite{Qiu2015} with a value close to $-40$~meV. By
comparing the PL spectra measured at zero magnetic field for the
different materials presented in this study (Fig.~\ref{fig:pl}),
the low temperature PL spectrum of MoS$_2$ resembles more the one
observed for WS$_2$ and WSe$_2$ than the one of MoSe$_2$ or of
MoTe$_2$~\cite{Ignacio2015}. In similarity to the low temperature
PL spectra of darkish monolayers, the spectrum of MoS$_2$ also
displays a significantly broad emission band at energies lower
than that of the A exciton [see Fig.~\ref{fig:pl}(d)]. The
observation of either a well defined two peaks PL spectrum arising
from the A and T excitons or an additional broad band associated
with localized/bound excitons~\cite{mak2012,cadizacid}, appears to
be characteristic of the two families of bright or darkish S-TMD
monolayers. The presence of emission due to localized/bound
excitons in the low temperature $B=0$ PL of darkish monolayers and
not in the bright ones could be due to the appearance of
long-lived reservoir of dark excitons in the former systems, which
then effectively diffuse and/or relax towards other possible
radiative centers. Similar relaxation processes can be largely
suppressed in bright monolayers, as the ground state excitons in
these systems already represent the effective recombination
channel. Following this logic, the ground state exciton should be
dark in the MoS$_2$ monolayer.

The darkish character of the MoS$_2$ monolayer is confirmed by our
magneto-PL study as indeed the in-plane magnetic field has a
dramatic effect on the PL spectrum of this monolayer. The relative
intensity spectrum displayed in the lower panel of
Fig.~\ref{fig:pl}(d) shows a rather single but broad peak,
labelled D, which is centered at about $97$~meV below the A
exciton line of MoS$_2$. Even though all observed PL peaks are
much broader in our MoS$_2$ monolayer than in other studied
materials, we have performed the same analysis as for the other
materials. Similarly to the case of WS$_2$ and WSe$_2$ monolayers,
the shape (width and center position) of the relative spectrum of
the MoS$_2$ monolayer remain field independent but its amplitude
increases quadratically, $\sim \alpha B^2$, with the magnetic
field [see Fig.~\ref{fig:dark}]. This result confirm placing the
MoS$_2$ in the family of darkish S-TMD with a dark-bright exciton
splitting twice bigger than that found in WS$_2$ or in WSe$_2$.

\section{Conclusions}

To conclude, we have presented the experimental investigations
supported by the theoretical consideration of the effect of
brightening of dark excitons in S-TMD monolayers induced by the
application of a magnetic field in the direction along the plane
of the layer. Field induced emission due to dark excitons can be
observed at low temperatures in S-TMD monolayers for which the
dark excitons are lower in energy than the bright excitons.
Emission intensities of dark excitons grow quadratically with the
strength of the in-plane magnetic field. These results lead us to
establish the WS$_2$, WSe$_2$ and MoS$_2$ monolayer as darkish
materials, $i.e.$, the direct bandgap systems but with a dark
excitonic ground state, and monolayers of MoSe$_2$ as a bright
materials with a bright exciton ground state. The bright-dark
exciton splitting is found to be of about 50~meV in WS$_2$ and
WSe$_2$ monolayers in fair agreement with theoretical
expectations~\cite{liu,kormanyos,Echeverry2016}, but its value
derived for the MoS$_2$ monolayer is surprisingly
large~\cite{Qiu2015}. The characteristic doublet structure of dark
excitons has been observed for WS$_2$, WSe$_2$ monolayers, along
the lines of the recent theoretical proposal~\cite{Slobo2016}.
Different ordering of the spin-orbit split subbands in the
conduction band for two, bright and darkish TMD families, is also
speculated to be reflected in the $B=0$ low temperature PL
spectra: bright monolayers show a simple emission due to exciton
and trions, the darkish ones display an additional broad/multipeak
emission band due to localised/bound excitons.

\section{Acknowledgements}

The work has been supported by the European Research Council (MOMB
project no. 320590), the EC Graphene Flagship project (no.
604391), the National Science Center (grant no.
DEC-2013/10/M/ST3/00791) and the Nanofab facility of the Institut
N\'eel, CNRS UGA.

\bibliography{biblio_dark}

\begin{thebibliography}{51}%
\makeatletter
\providecommand \@ifxundefined [1]{%
 \@ifx{#1\undefined}
}%
\providecommand \@ifnum [1]{%
 \ifnum #1\expandafter \@firstoftwo
 \else \expandafter \@secondoftwo
 \fi
}%
\providecommand \@ifx [1]{%
 \ifx #1\expandafter \@firstoftwo
 \else \expandafter \@secondoftwo
 \fi
}%
\providecommand \natexlab [1]{#1}%
\providecommand \enquote  [1]{``#1''}%
\providecommand \bibnamefont  [1]{#1}%
\providecommand \bibfnamefont [1]{#1}%
\providecommand \citenamefont [1]{#1}%
\providecommand \href@noop [0]{\@secondoftwo}%
\providecommand \href [0]{\begingroup \@sanitize@url \@href}%
\providecommand \@href[1]{\@@startlink{#1}\@@href}%
\providecommand \@@href[1]{\endgroup#1\@@endlink}%
\providecommand \@sanitize@url [0]{\catcode `\\12\catcode `\$12\catcode
  `\&12\catcode `\#12\catcode `\^12\catcode `\_12\catcode `\%12\relax}%
\providecommand \@@startlink[1]{}%
\providecommand \@@endlink[0]{}%
\providecommand \url  [0]{\begingroup\@sanitize@url \@url }%
\providecommand \@url [1]{\endgroup\@href {#1}{\urlprefix }}%
\providecommand \urlprefix  [0]{URL }%
\providecommand \Eprint [0]{\href }%
\providecommand \doibase [0]{http://dx.doi.org/}%
\providecommand \selectlanguage [0]{\@gobble}%
\providecommand \bibinfo  [0]{\@secondoftwo}%
\providecommand \bibfield  [0]{\@secondoftwo}%
\providecommand \translation [1]{[#1]}%
\providecommand \BibitemOpen [0]{}%
\providecommand \bibitemStop [0]{}%
\providecommand \bibitemNoStop [0]{.\EOS\space}%
\providecommand \EOS [0]{\spacefactor3000\relax}%
\providecommand \BibitemShut  [1]{\csname bibitem#1\endcsname}%
\let\auto@bib@innerbib\@empty
\bibitem [{\citenamefont {Mak}\ \emph {et~al.}(2010)\citenamefont {Mak},
  \citenamefont {Lee}, \citenamefont {Hone}, \citenamefont {Shan},\ and\
  \citenamefont {Heinz}}]{mak2010}%
  \BibitemOpen
  \bibfield  {author} {\bibinfo {author} {\bibfnamefont {K.~F.}\ \bibnamefont
  {Mak}}, \bibinfo {author} {\bibfnamefont {C.}~\bibnamefont {Lee}}, \bibinfo
  {author} {\bibfnamefont {J.}~\bibnamefont {Hone}}, \bibinfo {author}
  {\bibfnamefont {J.}~\bibnamefont {Shan}}, \ and\ \bibinfo {author}
  {\bibfnamefont {T.~F.}\ \bibnamefont {Heinz}},\ }\href {\doibase
  10.1103/PhysRevLett.105.136805} {\bibfield  {journal} {\bibinfo  {journal}
  {Phys. Rev. Lett.}\ }\textbf {\bibinfo {volume} {105}},\ \bibinfo {pages}
  {136805} (\bibinfo {year} {2010})}\BibitemShut {NoStop}%
\bibitem [{\citenamefont {Ugeda}\ \emph {et~al.}(2014)\citenamefont {Ugeda},
  \citenamefont {Bradley}, \citenamefont {Shi}, \citenamefont {da~Jornada},
  \citenamefont {Zhang}, \citenamefont {Qiu}, \citenamefont {Ruan},
  \citenamefont {Mo}, \citenamefont {Hussain}, \citenamefont {Shen},
  \citenamefont {Wang}, \citenamefont {Louie},\ and\ \citenamefont
  {Crommie}}]{ugeda2014}%
  \BibitemOpen
  \bibfield  {author} {\bibinfo {author} {\bibfnamefont {M.~M.}\ \bibnamefont
  {Ugeda}}, \bibinfo {author} {\bibfnamefont {A.~J.}\ \bibnamefont {Bradley}},
  \bibinfo {author} {\bibfnamefont {S.-F.}\ \bibnamefont {Shi}}, \bibinfo
  {author} {\bibfnamefont {F.~H.}\ \bibnamefont {da~Jornada}}, \bibinfo
  {author} {\bibfnamefont {Y.}~\bibnamefont {Zhang}}, \bibinfo {author}
  {\bibfnamefont {D.~Y.}\ \bibnamefont {Qiu}}, \bibinfo {author} {\bibfnamefont
  {W.}~\bibnamefont {Ruan}}, \bibinfo {author} {\bibfnamefont {S.-K.}\
  \bibnamefont {Mo}}, \bibinfo {author} {\bibfnamefont {Z.}~\bibnamefont
  {Hussain}}, \bibinfo {author} {\bibfnamefont {Z.-X.}\ \bibnamefont {Shen}},
  \bibinfo {author} {\bibfnamefont {F.}~\bibnamefont {Wang}}, \bibinfo {author}
  {\bibfnamefont {S.~G.}\ \bibnamefont {Louie}}, \ and\ \bibinfo {author}
  {\bibfnamefont {M.~F.}\ \bibnamefont {Crommie}},\ }\href {\doibase
  10.1038/nmat4061} {\bibfield  {journal} {\bibinfo  {journal} {Nat. Mater.}\
  }\textbf {\bibinfo {volume} {13}},\ \bibinfo {pages} {1091} (\bibinfo {year}
  {2014})}\BibitemShut {NoStop}%
\bibitem [{\citenamefont {Ye}\ \emph {et~al.}(2014)\citenamefont {Ye},
  \citenamefont {Cao}, \citenamefont {O'Brien}, \citenamefont {Zhu},
  \citenamefont {Yin}, \citenamefont {Wang}, \citenamefont {Louie},\ and\
  \citenamefont {Zhang}}]{ye}%
  \BibitemOpen
  \bibfield  {author} {\bibinfo {author} {\bibfnamefont {Z.}~\bibnamefont
  {Ye}}, \bibinfo {author} {\bibfnamefont {T.}~\bibnamefont {Cao}}, \bibinfo
  {author} {\bibfnamefont {K.}~\bibnamefont {O'Brien}}, \bibinfo {author}
  {\bibfnamefont {H.}~\bibnamefont {Zhu}}, \bibinfo {author} {\bibfnamefont
  {X.}~\bibnamefont {Yin}}, \bibinfo {author} {\bibfnamefont {Y.}~\bibnamefont
  {Wang}}, \bibinfo {author} {\bibfnamefont {S.~G.}\ \bibnamefont {Louie}}, \
  and\ \bibinfo {author} {\bibfnamefont {X.}~\bibnamefont {Zhang}},\ }\href
  {\doibase 10.1038/nature13734} {\bibfield  {journal} {\bibinfo  {journal}
  {Nature}\ }\textbf {\bibinfo {volume} {513}},\ \bibinfo {pages} {214}
  (\bibinfo {year} {2014})}\BibitemShut {NoStop}%
\bibitem [{\citenamefont {Chernikov}\ \emph {et~al.}(2015)\citenamefont
  {Chernikov}, \citenamefont {van~der Zande}, \citenamefont {Hill},
  \citenamefont {Rigosi}, \citenamefont {Velauthapillai}, \citenamefont
  {Hone},\ and\ \citenamefont {Heinz}}]{chernikov2015}%
  \BibitemOpen
  \bibfield  {author} {\bibinfo {author} {\bibfnamefont {A.}~\bibnamefont
  {Chernikov}}, \bibinfo {author} {\bibfnamefont {A.~M.}\ \bibnamefont {van~der
  Zande}}, \bibinfo {author} {\bibfnamefont {H.~M.}\ \bibnamefont {Hill}},
  \bibinfo {author} {\bibfnamefont {A.~F.}\ \bibnamefont {Rigosi}}, \bibinfo
  {author} {\bibfnamefont {A.}~\bibnamefont {Velauthapillai}}, \bibinfo
  {author} {\bibfnamefont {J.}~\bibnamefont {Hone}}, \ and\ \bibinfo {author}
  {\bibfnamefont {T.~F.}\ \bibnamefont {Heinz}},\ }\href {\doibase
  10.1103/PhysRevLett.115.126802} {\bibfield  {journal} {\bibinfo  {journal}
  {Phys. Rev. Lett.}\ }\textbf {\bibinfo {volume} {115}},\ \bibinfo {pages}
  {126802} (\bibinfo {year} {2015})}\BibitemShut {NoStop}%
\bibitem [{\citenamefont {Xiao}\ \emph {et~al.}(2012)\citenamefont {Xiao},
  \citenamefont {Liu}, \citenamefont {Feng}, \citenamefont {Xu},\ and\
  \citenamefont {Yao}}]{xiao}%
  \BibitemOpen
  \bibfield  {author} {\bibinfo {author} {\bibfnamefont {D.}~\bibnamefont
  {Xiao}}, \bibinfo {author} {\bibfnamefont {G.-B.}\ \bibnamefont {Liu}},
  \bibinfo {author} {\bibfnamefont {W.}~\bibnamefont {Feng}}, \bibinfo {author}
  {\bibfnamefont {X.}~\bibnamefont {Xu}}, \ and\ \bibinfo {author}
  {\bibfnamefont {W.}~\bibnamefont {Yao}},\ }\href {\doibase
  10.1103/PhysRevLett.108.196802} {\bibfield  {journal} {\bibinfo  {journal}
  {Phys. Rev. Lett.}\ }\textbf {\bibinfo {volume} {108}},\ \bibinfo {pages}
  {196802} (\bibinfo {year} {2012})}\BibitemShut {NoStop}%
\bibitem [{\citenamefont {Mak}\ \emph {et~al.}(2012)\citenamefont {Mak},
  \citenamefont {He}, \citenamefont {Shan},\ and\ \citenamefont
  {Heinz}}]{mak2012}%
  \BibitemOpen
  \bibfield  {author} {\bibinfo {author} {\bibfnamefont {K.~F.}\ \bibnamefont
  {Mak}}, \bibinfo {author} {\bibfnamefont {K.}~\bibnamefont {He}}, \bibinfo
  {author} {\bibfnamefont {J.}~\bibnamefont {Shan}}, \ and\ \bibinfo {author}
  {\bibfnamefont {T.~F.}\ \bibnamefont {Heinz}},\ }\href {\doibase
  10.1038/nnano.2012.96} {\bibfield  {journal} {\bibinfo  {journal} {Nat.
  Nanotechnol.}\ }\textbf {\bibinfo {volume} {7}},\ \bibinfo {pages} {494}
  (\bibinfo {year} {2012})}\BibitemShut {NoStop}%
\bibitem [{\citenamefont {Cao}\ \emph {et~al.}(2012)\citenamefont {Cao},
  \citenamefont {Wang}, \citenamefont {Han}, \citenamefont {Ye}, \citenamefont
  {Zhu}, \citenamefont {Shi}, \citenamefont {Niu}, \citenamefont {Tan},
  \citenamefont {Wang}, \citenamefont {Liu},\ and\ \citenamefont {Feng}}]{cao}%
  \BibitemOpen
  \bibfield  {author} {\bibinfo {author} {\bibfnamefont {T.}~\bibnamefont
  {Cao}}, \bibinfo {author} {\bibfnamefont {G.}~\bibnamefont {Wang}}, \bibinfo
  {author} {\bibfnamefont {W.}~\bibnamefont {Han}}, \bibinfo {author}
  {\bibfnamefont {H.}~\bibnamefont {Ye}}, \bibinfo {author} {\bibfnamefont
  {C.}~\bibnamefont {Zhu}}, \bibinfo {author} {\bibfnamefont {J.}~\bibnamefont
  {Shi}}, \bibinfo {author} {\bibfnamefont {Q.}~\bibnamefont {Niu}}, \bibinfo
  {author} {\bibfnamefont {P.}~\bibnamefont {Tan}}, \bibinfo {author}
  {\bibfnamefont {E.}~\bibnamefont {Wang}}, \bibinfo {author} {\bibfnamefont
  {B.}~\bibnamefont {Liu}}, \ and\ \bibinfo {author} {\bibfnamefont
  {J.}~\bibnamefont {Feng}},\ }\href {\doibase 10.1038/ncomms1882} {\bibfield
  {journal} {\bibinfo  {journal} {Nat. Commun.}\ }\textbf {\bibinfo {volume}
  {3}},\ \bibinfo {pages} {887} (\bibinfo {year} {2012})}\BibitemShut {NoStop}%
\bibitem [{\citenamefont {Zeng}\ \emph {et~al.}(2012)\citenamefont {Zeng},
  \citenamefont {Dai}, \citenamefont {Yao}, \citenamefont {Xiao},\ and\
  \citenamefont {Cui}}]{zeng2012}%
  \BibitemOpen
  \bibfield  {author} {\bibinfo {author} {\bibfnamefont {H.}~\bibnamefont
  {Zeng}}, \bibinfo {author} {\bibfnamefont {J.}~\bibnamefont {Dai}}, \bibinfo
  {author} {\bibfnamefont {W.}~\bibnamefont {Yao}}, \bibinfo {author}
  {\bibfnamefont {D.}~\bibnamefont {Xiao}}, \ and\ \bibinfo {author}
  {\bibfnamefont {X.}~\bibnamefont {Cui}},\ }\href {\doibase
  10.1038/nnano.2012.95} {\bibfield  {journal} {\bibinfo  {journal} {Nat.
  Nanotechnol.}\ }\textbf {\bibinfo {volume} {7}},\ \bibinfo {pages} {490}
  (\bibinfo {year} {2012})}\BibitemShut {NoStop}%
\bibitem [{\citenamefont {Jones}\ \emph {et~al.}(2013)\citenamefont {Jones},
  \citenamefont {Yu}, \citenamefont {Ghimire}, \citenamefont {Wu},
  \citenamefont {Aivazian}, \citenamefont {Ross}, \citenamefont {Zhao},
  \citenamefont {Yan}, \citenamefont {Mandrus}, \citenamefont {Xiao},
  \citenamefont {Yao},\ and\ \citenamefont {Xu}}]{jones}%
  \BibitemOpen
  \bibfield  {author} {\bibinfo {author} {\bibfnamefont {A.~M.}\ \bibnamefont
  {Jones}}, \bibinfo {author} {\bibfnamefont {H.}~\bibnamefont {Yu}}, \bibinfo
  {author} {\bibfnamefont {N.~J.}\ \bibnamefont {Ghimire}}, \bibinfo {author}
  {\bibfnamefont {S.}~\bibnamefont {Wu}}, \bibinfo {author} {\bibfnamefont
  {G.}~\bibnamefont {Aivazian}}, \bibinfo {author} {\bibfnamefont {J.~S.}\
  \bibnamefont {Ross}}, \bibinfo {author} {\bibfnamefont {B.}~\bibnamefont
  {Zhao}}, \bibinfo {author} {\bibfnamefont {J.}~\bibnamefont {Yan}}, \bibinfo
  {author} {\bibfnamefont {D.~G.}\ \bibnamefont {Mandrus}}, \bibinfo {author}
  {\bibfnamefont {D.}~\bibnamefont {Xiao}}, \bibinfo {author} {\bibfnamefont
  {W.}~\bibnamefont {Yao}}, \ and\ \bibinfo {author} {\bibfnamefont
  {X.}~\bibnamefont {Xu}},\ }\href {\doibase 10.1038/nnano.2013.151} {\bibfield
   {journal} {\bibinfo  {journal} {Nat. Nanotechnol.}\ }\textbf {\bibinfo
  {volume} {8}},\ \bibinfo {pages} {634} (\bibinfo {year} {2013})}\BibitemShut
  {NoStop}%
\bibitem [{\citenamefont {Wang}\ \emph {et~al.}(2016)\citenamefont {Wang},
  \citenamefont {Marie}, \citenamefont {Liu}, \citenamefont {Amand},
  \citenamefont {Robert}, \citenamefont {Cadiz}, \citenamefont {Renucci},\ and\
  \citenamefont {Urbaszek}}]{wang2016}%
  \BibitemOpen
  \bibfield  {author} {\bibinfo {author} {\bibfnamefont {G.}~\bibnamefont
  {Wang}}, \bibinfo {author} {\bibfnamefont {X.}~\bibnamefont {Marie}},
  \bibinfo {author} {\bibfnamefont {B.~L.}\ \bibnamefont {Liu}}, \bibinfo
  {author} {\bibfnamefont {T.}~\bibnamefont {Amand}}, \bibinfo {author}
  {\bibfnamefont {C.}~\bibnamefont {Robert}}, \bibinfo {author} {\bibfnamefont
  {F.}~\bibnamefont {Cadiz}}, \bibinfo {author} {\bibfnamefont
  {P.}~\bibnamefont {Renucci}}, \ and\ \bibinfo {author} {\bibfnamefont
  {B.}~\bibnamefont {Urbaszek}},\ }\href {\doibase
  10.1103/PhysRevLett.117.187401} {\bibfield  {journal} {\bibinfo  {journal}
  {Phys. Rev. Lett.}\ }\textbf {\bibinfo {volume} {117}},\ \bibinfo {pages}
  {187401} (\bibinfo {year} {2016})}\BibitemShut {NoStop}%
\bibitem [{\citenamefont {Baugher}\ \emph {et~al.}(2014)\citenamefont
  {Baugher}, \citenamefont {Churchill}, \citenamefont {Yang},\ and\
  \citenamefont {Jarillo-Herrero}}]{britton}%
  \BibitemOpen
  \bibfield  {author} {\bibinfo {author} {\bibfnamefont {B.~W.~H.}\
  \bibnamefont {Baugher}}, \bibinfo {author} {\bibfnamefont {H.~O.~H.}\
  \bibnamefont {Churchill}}, \bibinfo {author} {\bibfnamefont {Y.}~\bibnamefont
  {Yang}}, \ and\ \bibinfo {author} {\bibfnamefont {P.}~\bibnamefont
  {Jarillo-Herrero}},\ }\href {\doibase 10.1038/nnano.2014.25} {\bibfield
  {journal} {\bibinfo  {journal} {Nat. Nanotechnol.}\ }\textbf {\bibinfo
  {volume} {9}},\ \bibinfo {pages} {262} (\bibinfo {year} {2014})}\BibitemShut
  {NoStop}%
\bibitem [{\citenamefont {Zhang}\ \emph {et~al.}(2014)\citenamefont {Zhang},
  \citenamefont {Chang}, \citenamefont {Zhou}, \citenamefont {Cui},
  \citenamefont {Yan}, \citenamefont {Liu}, \citenamefont {Schmitt},
  \citenamefont {Lee}, \citenamefont {Moore}, \citenamefont {Chen},
  \citenamefont {Lin}, \citenamefont {Jeng}, \citenamefont {Mo}, \citenamefont
  {Hussain}, \citenamefont {Bansil},\ and\ \citenamefont {Shen}}]{zhang}%
  \BibitemOpen
  \bibfield  {author} {\bibinfo {author} {\bibfnamefont {Y.}~\bibnamefont
  {Zhang}}, \bibinfo {author} {\bibfnamefont {T.-R.}\ \bibnamefont {Chang}},
  \bibinfo {author} {\bibfnamefont {B.}~\bibnamefont {Zhou}}, \bibinfo {author}
  {\bibfnamefont {Y.-T.}\ \bibnamefont {Cui}}, \bibinfo {author} {\bibfnamefont
  {H.}~\bibnamefont {Yan}}, \bibinfo {author} {\bibfnamefont {Z.}~\bibnamefont
  {Liu}}, \bibinfo {author} {\bibfnamefont {F.}~\bibnamefont {Schmitt}},
  \bibinfo {author} {\bibfnamefont {J.}~\bibnamefont {Lee}}, \bibinfo {author}
  {\bibfnamefont {R.}~\bibnamefont {Moore}}, \bibinfo {author} {\bibfnamefont
  {Y.}~\bibnamefont {Chen}}, \bibinfo {author} {\bibfnamefont {H.}~\bibnamefont
  {Lin}}, \bibinfo {author} {\bibfnamefont {H.-T.}\ \bibnamefont {Jeng}},
  \bibinfo {author} {\bibfnamefont {S.-K.}\ \bibnamefont {Mo}}, \bibinfo
  {author} {\bibfnamefont {Z.}~\bibnamefont {Hussain}}, \bibinfo {author}
  {\bibfnamefont {A.}~\bibnamefont {Bansil}}, \ and\ \bibinfo {author}
  {\bibfnamefont {Z.-X.}\ \bibnamefont {Shen}},\ }\href {\doibase
  10.1038/nnano.2013.277} {\bibfield  {journal} {\bibinfo  {journal} {Nat.
  Nanotechnol.}\ }\textbf {\bibinfo {volume} {9}},\ \bibinfo {pages} {111}
  (\bibinfo {year} {2014})}\BibitemShut {NoStop}%
\bibitem [{\citenamefont {Riley}\ \emph {et~al.}(2014)\citenamefont {Riley},
  \citenamefont {Mazzola}, \citenamefont {Dendzik}, \citenamefont {Michiardi},
  \citenamefont {Takayama}, \citenamefont {Bawden}, \citenamefont
  {Graner{\o}d}, \citenamefont {Leandersson}, \citenamefont {Balasubramanian},
  \citenamefont {Hoesch}, \citenamefont {Kim}, \citenamefont {Takagi},
  \citenamefont {Meevasana}, \citenamefont {Hofmann}, \citenamefont {Bahramy},
  \citenamefont {Wells},\ and\ \citenamefont {King}}]{riley2014}%
  \BibitemOpen
  \bibfield  {author} {\bibinfo {author} {\bibfnamefont {J.~M.}\ \bibnamefont
  {Riley}}, \bibinfo {author} {\bibfnamefont {F.}~\bibnamefont {Mazzola}},
  \bibinfo {author} {\bibfnamefont {M.}~\bibnamefont {Dendzik}}, \bibinfo
  {author} {\bibfnamefont {M.}~\bibnamefont {Michiardi}}, \bibinfo {author}
  {\bibfnamefont {T.}~\bibnamefont {Takayama}}, \bibinfo {author}
  {\bibfnamefont {L.}~\bibnamefont {Bawden}}, \bibinfo {author} {\bibfnamefont
  {C.}~\bibnamefont {Graner{\o}d}}, \bibinfo {author} {\bibfnamefont
  {M.}~\bibnamefont {Leandersson}}, \bibinfo {author} {\bibfnamefont
  {T.}~\bibnamefont {Balasubramanian}}, \bibinfo {author} {\bibfnamefont
  {M.}~\bibnamefont {Hoesch}}, \bibinfo {author} {\bibfnamefont {T.~K.}\
  \bibnamefont {Kim}}, \bibinfo {author} {\bibfnamefont {H.}~\bibnamefont
  {Takagi}}, \bibinfo {author} {\bibfnamefont {W.}~\bibnamefont {Meevasana}},
  \bibinfo {author} {\bibfnamefont {P.}~\bibnamefont {Hofmann}}, \bibinfo
  {author} {\bibfnamefont {M.~S.}\ \bibnamefont {Bahramy}}, \bibinfo {author}
  {\bibfnamefont {J.~W.}\ \bibnamefont {Wells}}, \ and\ \bibinfo {author}
  {\bibfnamefont {P.~D.~C.}\ \bibnamefont {King}},\ }\href {\doibase
  10.1038/nphys3105} {\bibfield  {journal} {\bibinfo  {journal} {Nat. Phys.}\
  }\textbf {\bibinfo {volume} {10}},\ \bibinfo {pages} {835} (\bibinfo {year}
  {2014})}\BibitemShut {NoStop}%
\bibitem [{\citenamefont {Ross}\ \emph {et~al.}(2013)\citenamefont {Ross},
  \citenamefont {Wu}, \citenamefont {Yu}, \citenamefont {Ghimire},
  \citenamefont {Jones}, \citenamefont {Aivazian}, \citenamefont {Yan},
  \citenamefont {Mandrus}, \citenamefont {Xiao}, \citenamefont {Yao},\ and\
  \citenamefont {Xu}}]{ross}%
  \BibitemOpen
  \bibfield  {author} {\bibinfo {author} {\bibfnamefont {J.~S.}\ \bibnamefont
  {Ross}}, \bibinfo {author} {\bibfnamefont {S.}~\bibnamefont {Wu}}, \bibinfo
  {author} {\bibfnamefont {H.}~\bibnamefont {Yu}}, \bibinfo {author}
  {\bibfnamefont {N.~J.}\ \bibnamefont {Ghimire}}, \bibinfo {author}
  {\bibfnamefont {A.~M.}\ \bibnamefont {Jones}}, \bibinfo {author}
  {\bibfnamefont {G.}~\bibnamefont {Aivazian}}, \bibinfo {author}
  {\bibfnamefont {J.}~\bibnamefont {Yan}}, \bibinfo {author} {\bibfnamefont
  {D.~G.}\ \bibnamefont {Mandrus}}, \bibinfo {author} {\bibfnamefont
  {D.}~\bibnamefont {Xiao}}, \bibinfo {author} {\bibfnamefont {W.}~\bibnamefont
  {Yao}}, \ and\ \bibinfo {author} {\bibfnamefont {X.}~\bibnamefont {Xu}},\
  }\href {\doibase 10.1038/ncomms2498} {\bibfield  {journal} {\bibinfo
  {journal} {Nat. Commun.}\ }\textbf {\bibinfo {volume} {4}},\ \bibinfo {pages}
  {1474} (\bibinfo {year} {2013})}\BibitemShut {NoStop}%
\bibitem [{\citenamefont {Chernikov}\ \emph {et~al.}(2014)\citenamefont
  {Chernikov}, \citenamefont {Berkelbach}, \citenamefont {Hill}, \citenamefont
  {Rigosi}, \citenamefont {Li}, \citenamefont {Aslan}, \citenamefont
  {Reichman}, \citenamefont {Hybertsen},\ and\ \citenamefont
  {Heinz}}]{chernikov}%
  \BibitemOpen
  \bibfield  {author} {\bibinfo {author} {\bibfnamefont {A.}~\bibnamefont
  {Chernikov}}, \bibinfo {author} {\bibfnamefont {T.~C.}\ \bibnamefont
  {Berkelbach}}, \bibinfo {author} {\bibfnamefont {H.~M.}\ \bibnamefont
  {Hill}}, \bibinfo {author} {\bibfnamefont {A.}~\bibnamefont {Rigosi}},
  \bibinfo {author} {\bibfnamefont {Y.}~\bibnamefont {Li}}, \bibinfo {author}
  {\bibfnamefont {O.~B.}\ \bibnamefont {Aslan}}, \bibinfo {author}
  {\bibfnamefont {D.~R.}\ \bibnamefont {Reichman}}, \bibinfo {author}
  {\bibfnamefont {M.~S.}\ \bibnamefont {Hybertsen}}, \ and\ \bibinfo {author}
  {\bibfnamefont {T.~F.}\ \bibnamefont {Heinz}},\ }\href {\doibase
  10.1103/PhysRevLett.113.076802} {\bibfield  {journal} {\bibinfo  {journal}
  {Phys. Rev. Lett.}\ }\textbf {\bibinfo {volume} {113}},\ \bibinfo {pages}
  {076802} (\bibinfo {year} {2014})}\BibitemShut {NoStop}%
\bibitem [{\citenamefont {Zhao}\ \emph {et~al.}(2012)\citenamefont {Zhao},
  \citenamefont {Ghorannevis}, \citenamefont {Chu}, \citenamefont {Toh},
  \citenamefont {Kloc}, \citenamefont {Tan},\ and\ \citenamefont
  {Eda}}]{zhao2012}%
  \BibitemOpen
  \bibfield  {author} {\bibinfo {author} {\bibfnamefont {W.}~\bibnamefont
  {Zhao}}, \bibinfo {author} {\bibfnamefont {Z.}~\bibnamefont {Ghorannevis}},
  \bibinfo {author} {\bibfnamefont {L.}~\bibnamefont {Chu}}, \bibinfo {author}
  {\bibfnamefont {M.}~\bibnamefont {Toh}}, \bibinfo {author} {\bibfnamefont
  {C.}~\bibnamefont {Kloc}}, \bibinfo {author} {\bibfnamefont {P.-H.}\
  \bibnamefont {Tan}}, \ and\ \bibinfo {author} {\bibfnamefont
  {G.}~\bibnamefont {Eda}},\ }\href {\doibase 10.1021/nn305275h} {\bibfield
  {journal} {\bibinfo  {journal} {ACS Nano}\ }\textbf {\bibinfo {volume} {7}},\
  \bibinfo {pages} {791} (\bibinfo {year} {2012})}\BibitemShut {NoStop}%
\bibitem [{\citenamefont {Kozawa}\ \emph {et~al.}(2014)\citenamefont {Kozawa},
  \citenamefont {Kumar}, \citenamefont {Carvalho}, \citenamefont {Amara},
  \citenamefont {Zhao}, \citenamefont {Wang}, \citenamefont {Toh},
  \citenamefont {Ribeiro}, \citenamefont {Neto}, \citenamefont {Matsuda},\ and\
  \citenamefont {Eda}}]{kozawa}%
  \BibitemOpen
  \bibfield  {author} {\bibinfo {author} {\bibfnamefont {D.}~\bibnamefont
  {Kozawa}}, \bibinfo {author} {\bibfnamefont {R.}~\bibnamefont {Kumar}},
  \bibinfo {author} {\bibfnamefont {A.}~\bibnamefont {Carvalho}}, \bibinfo
  {author} {\bibfnamefont {K.~K.}\ \bibnamefont {Amara}}, \bibinfo {author}
  {\bibfnamefont {W.}~\bibnamefont {Zhao}}, \bibinfo {author} {\bibfnamefont
  {S.}~\bibnamefont {Wang}}, \bibinfo {author} {\bibfnamefont {M.}~\bibnamefont
  {Toh}}, \bibinfo {author} {\bibfnamefont {R.~M.}\ \bibnamefont {Ribeiro}},
  \bibinfo {author} {\bibfnamefont {A.~H.~C.}\ \bibnamefont {Neto}}, \bibinfo
  {author} {\bibfnamefont {K.}~\bibnamefont {Matsuda}}, \ and\ \bibinfo
  {author} {\bibfnamefont {G.}~\bibnamefont {Eda}},\ }\href {\doibase
  10.1038/ncomms5543} {\bibfield  {journal} {\bibinfo  {journal} {Nat.
  Commun.}\ }\textbf {\bibinfo {volume} {5}},\ \bibinfo {pages} {4543}
  (\bibinfo {year} {2014})}\BibitemShut {NoStop}%
\bibitem [{\citenamefont {Zeng}\ \emph {et~al.}(2013)\citenamefont {Zeng},
  \citenamefont {Liu}, \citenamefont {Dai}, \citenamefont {Yan}, \citenamefont
  {Zhu}, \citenamefont {He}, \citenamefont {Xie}, \citenamefont {Xu},
  \citenamefont {Chen}, \citenamefont {Yao},\ and\ \citenamefont {Cui}}]{zeng}%
  \BibitemOpen
  \bibfield  {author} {\bibinfo {author} {\bibfnamefont {H.}~\bibnamefont
  {Zeng}}, \bibinfo {author} {\bibfnamefont {G.-B.}\ \bibnamefont {Liu}},
  \bibinfo {author} {\bibfnamefont {J.}~\bibnamefont {Dai}}, \bibinfo {author}
  {\bibfnamefont {Y.}~\bibnamefont {Yan}}, \bibinfo {author} {\bibfnamefont
  {B.}~\bibnamefont {Zhu}}, \bibinfo {author} {\bibfnamefont {R.}~\bibnamefont
  {He}}, \bibinfo {author} {\bibfnamefont {L.}~\bibnamefont {Xie}}, \bibinfo
  {author} {\bibfnamefont {S.}~\bibnamefont {Xu}}, \bibinfo {author}
  {\bibfnamefont {X.}~\bibnamefont {Chen}}, \bibinfo {author} {\bibfnamefont
  {W.}~\bibnamefont {Yao}}, \ and\ \bibinfo {author} {\bibfnamefont
  {X.}~\bibnamefont {Cui}},\ }\href {\doibase 10.1038/srep01608} {\bibfield
  {journal} {\bibinfo  {journal} {Sci. Rep.}\ }\textbf {\bibinfo {volume}
  {3}},\ \bibinfo {pages} {1608} (\bibinfo {year} {2013})}\BibitemShut
  {NoStop}%
\bibitem [{\citenamefont {Li}\ \emph {et~al.}(2014{\natexlab{a}})\citenamefont
  {Li}, \citenamefont {Birdwell}, \citenamefont {Amani}, \citenamefont {Burke},
  \citenamefont {Ling}, \citenamefont {Lee}, \citenamefont {Liang},
  \citenamefont {Peng}, \citenamefont {Richter}, \citenamefont {Kong},
  \citenamefont {Gundlach},\ and\ \citenamefont {Nguyen}}]{liPRB}%
  \BibitemOpen
  \bibfield  {author} {\bibinfo {author} {\bibfnamefont {W.}~\bibnamefont
  {Li}}, \bibinfo {author} {\bibfnamefont {A.~G.}\ \bibnamefont {Birdwell}},
  \bibinfo {author} {\bibfnamefont {M.}~\bibnamefont {Amani}}, \bibinfo
  {author} {\bibfnamefont {R.~A.}\ \bibnamefont {Burke}}, \bibinfo {author}
  {\bibfnamefont {X.}~\bibnamefont {Ling}}, \bibinfo {author} {\bibfnamefont
  {Y.-H.}\ \bibnamefont {Lee}}, \bibinfo {author} {\bibfnamefont
  {X.}~\bibnamefont {Liang}}, \bibinfo {author} {\bibfnamefont
  {L.}~\bibnamefont {Peng}}, \bibinfo {author} {\bibfnamefont {C.~A.}\
  \bibnamefont {Richter}}, \bibinfo {author} {\bibfnamefont {J.}~\bibnamefont
  {Kong}}, \bibinfo {author} {\bibfnamefont {D.~J.}\ \bibnamefont {Gundlach}},
  \ and\ \bibinfo {author} {\bibfnamefont {N.~V.}\ \bibnamefont {Nguyen}},\
  }\href {\doibase 10.1103/PhysRevB.90.195434} {\bibfield  {journal} {\bibinfo
  {journal} {Phys. Rev. B}\ }\textbf {\bibinfo {volume} {90}},\ \bibinfo
  {pages} {195434} (\bibinfo {year} {2014}{\natexlab{a}})}\BibitemShut
  {NoStop}%
\bibitem [{\citenamefont {Zhu}\ \emph {et~al.}(2015)\citenamefont {Zhu},
  \citenamefont {Chen},\ and\ \citenamefont {Cui}}]{zhu}%
  \BibitemOpen
  \bibfield  {author} {\bibinfo {author} {\bibfnamefont {B.}~\bibnamefont
  {Zhu}}, \bibinfo {author} {\bibfnamefont {X.}~\bibnamefont {Chen}}, \ and\
  \bibinfo {author} {\bibfnamefont {X.}~\bibnamefont {Cui}},\ }\href {\doibase
  10.1038/srep09218} {\bibfield  {journal} {\bibinfo  {journal} {Sci. Rep.}\
  }\textbf {\bibinfo {volume} {5}},\ \bibinfo {pages} {9218} (\bibinfo {year}
  {2015})}\BibitemShut {NoStop}%
\bibitem [{\citenamefont {Klots}\ \emph {et~al.}(2014)\citenamefont {Klots},
  \citenamefont {Newaz}, \citenamefont {Wang}, \citenamefont {Prasai},
  \citenamefont {Krzyzanowska}, \citenamefont {Lin}, \citenamefont {Caudel},
  \citenamefont {Ghimire}, \citenamefont {Yan}, \citenamefont {Ivanov},
  \citenamefont {Velizhanin}, \citenamefont {Burger}, \citenamefont {Mandrus},
  \citenamefont {Tolk}, \citenamefont {Pantelides},\ and\ \citenamefont
  {Bolotin}}]{klots}%
  \BibitemOpen
  \bibfield  {author} {\bibinfo {author} {\bibfnamefont {A.~R.}\ \bibnamefont
  {Klots}}, \bibinfo {author} {\bibfnamefont {A.~K.~M.}\ \bibnamefont {Newaz}},
  \bibinfo {author} {\bibfnamefont {B.}~\bibnamefont {Wang}}, \bibinfo {author}
  {\bibfnamefont {D.}~\bibnamefont {Prasai}}, \bibinfo {author} {\bibfnamefont
  {H.}~\bibnamefont {Krzyzanowska}}, \bibinfo {author} {\bibfnamefont
  {J.}~\bibnamefont {Lin}}, \bibinfo {author} {\bibfnamefont {D.}~\bibnamefont
  {Caudel}}, \bibinfo {author} {\bibfnamefont {N.~J.}\ \bibnamefont {Ghimire}},
  \bibinfo {author} {\bibfnamefont {J.}~\bibnamefont {Yan}}, \bibinfo {author}
  {\bibfnamefont {B.~L.}\ \bibnamefont {Ivanov}}, \bibinfo {author}
  {\bibfnamefont {K.~A.}\ \bibnamefont {Velizhanin}}, \bibinfo {author}
  {\bibfnamefont {A.}~\bibnamefont {Burger}}, \bibinfo {author} {\bibfnamefont
  {D.~G.}\ \bibnamefont {Mandrus}}, \bibinfo {author} {\bibfnamefont {N.~H.}\
  \bibnamefont {Tolk}}, \bibinfo {author} {\bibfnamefont {S.~T.}\ \bibnamefont
  {Pantelides}}, \ and\ \bibinfo {author} {\bibfnamefont {K.~I.}\ \bibnamefont
  {Bolotin}},\ }\href {\doibase 10.1038/srep06608} {\bibfield  {journal}
  {\bibinfo  {journal} {Sci. Rep.}\ }\textbf {\bibinfo {volume} {4}},\ \bibinfo
  {pages} {6608} (\bibinfo {year} {2014})}\BibitemShut {NoStop}%
\bibitem [{\citenamefont {Hanbicki}\ \emph {et~al.}(2015)\citenamefont
  {Hanbicki}, \citenamefont {Currie}, \citenamefont {Kioseoglou}, \citenamefont
  {Friedman},\ and\ \citenamefont {Jonker}}]{hanbicki}%
  \BibitemOpen
  \bibfield  {author} {\bibinfo {author} {\bibfnamefont {A.}~\bibnamefont
  {Hanbicki}}, \bibinfo {author} {\bibfnamefont {M.}~\bibnamefont {Currie}},
  \bibinfo {author} {\bibfnamefont {G.}~\bibnamefont {Kioseoglou}}, \bibinfo
  {author} {\bibfnamefont {A.}~\bibnamefont {Friedman}}, \ and\ \bibinfo
  {author} {\bibfnamefont {B.}~\bibnamefont {Jonker}},\ }\href {\doibase
  http://dx.doi.org/10.1016/j.ssc.2014.11.005} {\bibfield  {journal} {\bibinfo
  {journal} {Sol. State Commun.}\ }\textbf {\bibinfo {volume} {203}},\ \bibinfo
  {pages} {16} (\bibinfo {year} {2015})}\BibitemShut {NoStop}%
\bibitem [{\citenamefont {Ko\ifmmode~\acute{s}\else \'{s}\fi{}mider}\ and\
  \citenamefont {Fern\'andez-Rossier}(2013)}]{Kosmider2013a}%
  \BibitemOpen
  \bibfield  {author} {\bibinfo {author} {\bibfnamefont {K.}~\bibnamefont
  {Ko\ifmmode~\acute{s}\else \'{s}\fi{}mider}}\ and\ \bibinfo {author}
  {\bibfnamefont {J.}~\bibnamefont {Fern\'andez-Rossier}},\ }\href {\doibase
  10.1103/PhysRevB.87.075451} {\bibfield  {journal} {\bibinfo  {journal} {Phys.
  Rev. B}\ }\textbf {\bibinfo {volume} {87}},\ \bibinfo {pages} {075451}
  (\bibinfo {year} {2013})}\BibitemShut {NoStop}%
\bibitem [{\citenamefont {Ko\ifmmode~\acute{s}\else \'{s}\fi{}mider}\ \emph
  {et~al.}(2013)\citenamefont {Ko\ifmmode~\acute{s}\else \'{s}\fi{}mider},
  \citenamefont {Gonz\'alez},\ and\ \citenamefont
  {Fern\'andez-Rossier}}]{Kosmider2013b}%
  \BibitemOpen
  \bibfield  {author} {\bibinfo {author} {\bibfnamefont {K.}~\bibnamefont
  {Ko\ifmmode~\acute{s}\else \'{s}\fi{}mider}}, \bibinfo {author}
  {\bibfnamefont {J.~W.}\ \bibnamefont {Gonz\'alez}}, \ and\ \bibinfo {author}
  {\bibfnamefont {J.}~\bibnamefont {Fern\'andez-Rossier}},\ }\href {\doibase
  10.1103/PhysRevB.88.245436} {\bibfield  {journal} {\bibinfo  {journal} {Phys.
  Rev. B}\ }\textbf {\bibinfo {volume} {88}},\ \bibinfo {pages} {245436}
  (\bibinfo {year} {2013})}\BibitemShut {NoStop}%
\bibitem [{\citenamefont {Liu}\ \emph {et~al.}(2013)\citenamefont {Liu},
  \citenamefont {Shan}, \citenamefont {Yao}, \citenamefont {Yao},\ and\
  \citenamefont {Xiao}}]{liu}%
  \BibitemOpen
  \bibfield  {author} {\bibinfo {author} {\bibfnamefont {G.-B.}\ \bibnamefont
  {Liu}}, \bibinfo {author} {\bibfnamefont {W.-Y.}\ \bibnamefont {Shan}},
  \bibinfo {author} {\bibfnamefont {Y.}~\bibnamefont {Yao}}, \bibinfo {author}
  {\bibfnamefont {W.}~\bibnamefont {Yao}}, \ and\ \bibinfo {author}
  {\bibfnamefont {D.}~\bibnamefont {Xiao}},\ }\href {\doibase
  10.1103/PhysRevB.88.085433} {\bibfield  {journal} {\bibinfo  {journal} {Phys.
  Rev. B}\ }\textbf {\bibinfo {volume} {88}},\ \bibinfo {pages} {085433}
  (\bibinfo {year} {2013})}\BibitemShut {NoStop}%
\bibitem [{\citenamefont {Korm\'anyos}\ \emph {et~al.}(2015)\citenamefont
  {Korm\'anyos}, \citenamefont {Burkard}, \citenamefont {Gmitra}, \citenamefont
  {Fabian}, \citenamefont {Z\'olyomi}, \citenamefont {Drummond},\ and\
  \citenamefont {Fal'ko}}]{kormanyos}%
  \BibitemOpen
  \bibfield  {author} {\bibinfo {author} {\bibfnamefont {A.}~\bibnamefont
  {Korm\'anyos}}, \bibinfo {author} {\bibfnamefont {G.}~\bibnamefont
  {Burkard}}, \bibinfo {author} {\bibfnamefont {M.}~\bibnamefont {Gmitra}},
  \bibinfo {author} {\bibfnamefont {J.}~\bibnamefont {Fabian}}, \bibinfo
  {author} {\bibfnamefont {V.}~\bibnamefont {Z\'olyomi}}, \bibinfo {author}
  {\bibfnamefont {N.~D.}\ \bibnamefont {Drummond}}, \ and\ \bibinfo {author}
  {\bibfnamefont {V.}~\bibnamefont {Fal'ko}},\ }\href
  {http://stacks.iop.org/2053-1583/2/i=2/a=022001} {\bibfield  {journal}
  {\bibinfo  {journal} {2D Materials}\ }\textbf {\bibinfo {volume} {2}},\
  \bibinfo {pages} {022001} (\bibinfo {year} {2015})}\BibitemShut {NoStop}%
\bibitem [{\citenamefont {Echeverry}\ \emph {et~al.}(2016)\citenamefont
  {Echeverry}, \citenamefont {Urbaszek}, \citenamefont {Amand}, \citenamefont
  {Marie},\ and\ \citenamefont {Gerber}}]{Echeverry2016}%
  \BibitemOpen
  \bibfield  {author} {\bibinfo {author} {\bibfnamefont {J.~P.}\ \bibnamefont
  {Echeverry}}, \bibinfo {author} {\bibfnamefont {B.}~\bibnamefont {Urbaszek}},
  \bibinfo {author} {\bibfnamefont {T.}~\bibnamefont {Amand}}, \bibinfo
  {author} {\bibfnamefont {X.}~\bibnamefont {Marie}}, \ and\ \bibinfo {author}
  {\bibfnamefont {I.~C.}\ \bibnamefont {Gerber}},\ }\href {\doibase
  10.1103/PhysRevB.93.121107} {\bibfield  {journal} {\bibinfo  {journal} {Phys.
  Rev. B}\ }\textbf {\bibinfo {volume} {93}},\ \bibinfo {pages} {121107}
  (\bibinfo {year} {2016})}\BibitemShut {NoStop}%
\bibitem [{\citenamefont {Qiu}\ \emph {et~al.}(2015)\citenamefont {Qiu},
  \citenamefont {Cao},\ and\ \citenamefont {Louie}}]{Qiu2015}%
  \BibitemOpen
  \bibfield  {author} {\bibinfo {author} {\bibfnamefont {D.~Y.}\ \bibnamefont
  {Qiu}}, \bibinfo {author} {\bibfnamefont {T.}~\bibnamefont {Cao}}, \ and\
  \bibinfo {author} {\bibfnamefont {S.~G.}\ \bibnamefont {Louie}},\ }\href
  {\doibase 10.1103/PhysRevLett.115.176801} {\bibfield  {journal} {\bibinfo
  {journal} {Phys. Rev. Lett.}\ }\textbf {\bibinfo {volume} {115}},\ \bibinfo
  {pages} {176801} (\bibinfo {year} {2015})}\BibitemShut {NoStop}%
\bibitem [{\citenamefont {Glazov}\ \emph {et~al.}(2014)\citenamefont {Glazov},
  \citenamefont {Amand}, \citenamefont {Marie}, \citenamefont {Lagarde},
  \citenamefont {Bouet},\ and\ \citenamefont {Urbaszek}}]{Glazov2014}%
  \BibitemOpen
  \bibfield  {author} {\bibinfo {author} {\bibfnamefont {M.~M.}\ \bibnamefont
  {Glazov}}, \bibinfo {author} {\bibfnamefont {T.}~\bibnamefont {Amand}},
  \bibinfo {author} {\bibfnamefont {X.}~\bibnamefont {Marie}}, \bibinfo
  {author} {\bibfnamefont {D.}~\bibnamefont {Lagarde}}, \bibinfo {author}
  {\bibfnamefont {L.}~\bibnamefont {Bouet}}, \ and\ \bibinfo {author}
  {\bibfnamefont {B.}~\bibnamefont {Urbaszek}},\ }\href {\doibase
  10.1103/PhysRevB.89.201302} {\bibfield  {journal} {\bibinfo  {journal} {Phys.
  Rev. B}\ }\textbf {\bibinfo {volume} {89}},\ \bibinfo {pages} {201302}
  (\bibinfo {year} {2014})}\BibitemShut {NoStop}%
\bibitem [{\citenamefont {Smole\ifmmode~\acute{n}\else \'{n}\fi{}ski}\ \emph
  {et~al.}(2016)\citenamefont {Smole\ifmmode~\acute{n}\else \'{n}\fi{}ski},
  \citenamefont {Goryca}, \citenamefont {Koperski}, \citenamefont {Faugeras},
  \citenamefont {Kazimierczuk}, \citenamefont {Bogucki}, \citenamefont
  {Nogajewski}, \citenamefont {Kossacki},\ and\ \citenamefont
  {Potemski}}]{smolenski}%
  \BibitemOpen
  \bibfield  {author} {\bibinfo {author} {\bibfnamefont {T.}~\bibnamefont
  {Smole\ifmmode~\acute{n}\else \'{n}\fi{}ski}}, \bibinfo {author}
  {\bibfnamefont {M.}~\bibnamefont {Goryca}}, \bibinfo {author} {\bibfnamefont
  {M.}~\bibnamefont {Koperski}}, \bibinfo {author} {\bibfnamefont
  {C.}~\bibnamefont {Faugeras}}, \bibinfo {author} {\bibfnamefont
  {T.}~\bibnamefont {Kazimierczuk}}, \bibinfo {author} {\bibfnamefont
  {A.}~\bibnamefont {Bogucki}}, \bibinfo {author} {\bibfnamefont
  {K.}~\bibnamefont {Nogajewski}}, \bibinfo {author} {\bibfnamefont
  {P.}~\bibnamefont {Kossacki}}, \ and\ \bibinfo {author} {\bibfnamefont
  {M.}~\bibnamefont {Potemski}},\ }\href {\doibase 10.1103/PhysRevX.6.021024}
  {\bibfield  {journal} {\bibinfo  {journal} {Phys. Rev. X}\ }\textbf {\bibinfo
  {volume} {6}},\ \bibinfo {pages} {021024} (\bibinfo {year}
  {2016})}\BibitemShut {NoStop}%
\bibitem [{\citenamefont {Arora}\ \emph
  {et~al.}(2015{\natexlab{a}})\citenamefont {Arora}, \citenamefont {Koperski},
  \citenamefont {Nogajewski}, \citenamefont {Marcus}, \citenamefont
  {Faugeras},\ and\ \citenamefont {Potemski}}]{arora}%
  \BibitemOpen
  \bibfield  {author} {\bibinfo {author} {\bibfnamefont {A.}~\bibnamefont
  {Arora}}, \bibinfo {author} {\bibfnamefont {M.}~\bibnamefont {Koperski}},
  \bibinfo {author} {\bibfnamefont {K.}~\bibnamefont {Nogajewski}}, \bibinfo
  {author} {\bibfnamefont {J.}~\bibnamefont {Marcus}}, \bibinfo {author}
  {\bibfnamefont {C.}~\bibnamefont {Faugeras}}, \ and\ \bibinfo {author}
  {\bibfnamefont {M.}~\bibnamefont {Potemski}},\ }\href {\doibase
  10.1039/C5NR01536G} {\bibfield  {journal} {\bibinfo  {journal} {Nanoscale}\
  }\textbf {\bibinfo {volume} {7}},\ \bibinfo {pages} {10421} (\bibinfo {year}
  {2015}{\natexlab{a}})}\BibitemShut {NoStop}%
\bibitem [{\citenamefont {Zhang}\ \emph {et~al.}(2015)\citenamefont {Zhang},
  \citenamefont {You}, \citenamefont {Zhao},\ and\ \citenamefont
  {Heinz}}]{Zhang2015a}%
  \BibitemOpen
  \bibfield  {author} {\bibinfo {author} {\bibfnamefont {X.-X.}\ \bibnamefont
  {Zhang}}, \bibinfo {author} {\bibfnamefont {Y.}~\bibnamefont {You}}, \bibinfo
  {author} {\bibfnamefont {S.~Y.~F.}\ \bibnamefont {Zhao}}, \ and\ \bibinfo
  {author} {\bibfnamefont {T.~F.}\ \bibnamefont {Heinz}},\ }\href {\doibase
  10.1103/PhysRevLett.115.257403} {\bibfield  {journal} {\bibinfo  {journal}
  {Phys. Rev. Lett.}\ }\textbf {\bibinfo {volume} {115}},\ \bibinfo {pages}
  {257403} (\bibinfo {year} {2015})}\BibitemShut {NoStop}%
\bibitem [{\citenamefont {Wang}\ \emph {et~al.}(2015)\citenamefont {Wang},
  \citenamefont {Robert}, \citenamefont {Suslu}, \citenamefont {Chen},
  \citenamefont {Yang}, \citenamefont {Alamdari}, \citenamefont {Gerber},
  \citenamefont {Amand}, \citenamefont {Marie}, \citenamefont {Tongay},\ and\
  \citenamefont {Urbaszek}}]{Wang2015a}%
  \BibitemOpen
  \bibfield  {author} {\bibinfo {author} {\bibfnamefont {G.}~\bibnamefont
  {Wang}}, \bibinfo {author} {\bibfnamefont {C.}~\bibnamefont {Robert}},
  \bibinfo {author} {\bibfnamefont {A.}~\bibnamefont {Suslu}}, \bibinfo
  {author} {\bibfnamefont {B.}~\bibnamefont {Chen}}, \bibinfo {author}
  {\bibfnamefont {S.}~\bibnamefont {Yang}}, \bibinfo {author} {\bibfnamefont
  {S.}~\bibnamefont {Alamdari}}, \bibinfo {author} {\bibfnamefont {I.~C.}\
  \bibnamefont {Gerber}}, \bibinfo {author} {\bibfnamefont {T.}~\bibnamefont
  {Amand}}, \bibinfo {author} {\bibfnamefont {X.}~\bibnamefont {Marie}},
  \bibinfo {author} {\bibfnamefont {S.}~\bibnamefont {Tongay}}, \ and\ \bibinfo
  {author} {\bibfnamefont {B.}~\bibnamefont {Urbaszek}},\ }\href {\doibase
  10.1038/ncomms10110} {\bibfield  {journal} {\bibinfo  {journal} {Nat.
  Commun.}\ }\textbf {\bibinfo {volume} {6}},\ \bibinfo {pages} {10110}
  (\bibinfo {year} {2015})}\BibitemShut {NoStop}%
\bibitem [{\citenamefont {Brandt}\ \emph {et~al.}(2007)\citenamefont {Brandt},
  \citenamefont {Fr\"ohlich}, \citenamefont {Sandfort}, \citenamefont {Bayer},
  \citenamefont {Stolz},\ and\ \citenamefont {Naka}}]{Brandt2007}%
  \BibitemOpen
  \bibfield  {author} {\bibinfo {author} {\bibfnamefont {J.}~\bibnamefont
  {Brandt}}, \bibinfo {author} {\bibfnamefont {D.}~\bibnamefont {Fr\"ohlich}},
  \bibinfo {author} {\bibfnamefont {C.}~\bibnamefont {Sandfort}}, \bibinfo
  {author} {\bibfnamefont {M.}~\bibnamefont {Bayer}}, \bibinfo {author}
  {\bibfnamefont {H.}~\bibnamefont {Stolz}}, \ and\ \bibinfo {author}
  {\bibfnamefont {N.}~\bibnamefont {Naka}},\ }\href {\doibase
  10.1103/PhysRevLett.99.217403} {\bibfield  {journal} {\bibinfo  {journal}
  {Phys. Rev. Lett.}\ }\textbf {\bibinfo {volume} {99}},\ \bibinfo {pages}
  {217403} (\bibinfo {year} {2007})}\BibitemShut {NoStop}%
\bibitem [{\citenamefont {Nirmal}\ \emph {et~al.}(1995)\citenamefont {Nirmal},
  \citenamefont {Norris}, \citenamefont {Kuno}, \citenamefont {Bawendi},
  \citenamefont {Efros},\ and\ \citenamefont {Rosen}}]{Nirmal1995}%
  \BibitemOpen
  \bibfield  {author} {\bibinfo {author} {\bibfnamefont {M.}~\bibnamefont
  {Nirmal}}, \bibinfo {author} {\bibfnamefont {D.~J.}\ \bibnamefont {Norris}},
  \bibinfo {author} {\bibfnamefont {M.}~\bibnamefont {Kuno}}, \bibinfo {author}
  {\bibfnamefont {M.~G.}\ \bibnamefont {Bawendi}}, \bibinfo {author}
  {\bibfnamefont {A.~L.}\ \bibnamefont {Efros}}, \ and\ \bibinfo {author}
  {\bibfnamefont {M.}~\bibnamefont {Rosen}},\ }\href {\doibase
  10.1103/PhysRevLett.75.3728} {\bibfield  {journal} {\bibinfo  {journal}
  {Phys. Rev. Lett.}\ }\textbf {\bibinfo {volume} {75}},\ \bibinfo {pages}
  {3728} (\bibinfo {year} {1995})}\BibitemShut {NoStop}%
\bibitem [{\citenamefont {Bayer}\ \emph {et~al.}(2000)\citenamefont {Bayer},
  \citenamefont {Stern}, \citenamefont {Kuther},\ and\ \citenamefont
  {Forchel}}]{Bayer2000}%
  \BibitemOpen
  \bibfield  {author} {\bibinfo {author} {\bibfnamefont {M.}~\bibnamefont
  {Bayer}}, \bibinfo {author} {\bibfnamefont {O.}~\bibnamefont {Stern}},
  \bibinfo {author} {\bibfnamefont {A.}~\bibnamefont {Kuther}}, \ and\ \bibinfo
  {author} {\bibfnamefont {A.}~\bibnamefont {Forchel}},\ }\href {\doibase
  10.1103/PhysRevB.61.7273} {\bibfield  {journal} {\bibinfo  {journal} {Phys.
  Rev. B}\ }\textbf {\bibinfo {volume} {61}},\ \bibinfo {pages} {7273}
  (\bibinfo {year} {2000})}\BibitemShut {NoStop}%
\bibitem [{\citenamefont {Zaric}\ \emph {et~al.}(2004)\citenamefont {Zaric},
  \citenamefont {Ostojic}, \citenamefont {Kono}, \citenamefont {Shaver},
  \citenamefont {Moore}, \citenamefont {Strano}, \citenamefont {Hauge},
  \citenamefont {Smalley},\ and\ \citenamefont {Wei}}]{Zaric2004}%
  \BibitemOpen
  \bibfield  {author} {\bibinfo {author} {\bibfnamefont {S.}~\bibnamefont
  {Zaric}}, \bibinfo {author} {\bibfnamefont {G.~N.}\ \bibnamefont {Ostojic}},
  \bibinfo {author} {\bibfnamefont {J.}~\bibnamefont {Kono}}, \bibinfo {author}
  {\bibfnamefont {J.}~\bibnamefont {Shaver}}, \bibinfo {author} {\bibfnamefont
  {V.~C.}\ \bibnamefont {Moore}}, \bibinfo {author} {\bibfnamefont {M.~S.}\
  \bibnamefont {Strano}}, \bibinfo {author} {\bibfnamefont {R.~H.}\
  \bibnamefont {Hauge}}, \bibinfo {author} {\bibfnamefont {R.~E.}\ \bibnamefont
  {Smalley}}, \ and\ \bibinfo {author} {\bibfnamefont {X.}~\bibnamefont
  {Wei}},\ }\href {\doibase 10.1126/science.1096524} {\bibfield  {journal}
  {\bibinfo  {journal} {Science}\ }\textbf {\bibinfo {volume} {304}},\ \bibinfo
  {pages} {1129} (\bibinfo {year} {2004})}\BibitemShut {NoStop}%
\bibitem [{\citenamefont {Srivastava}\ \emph {et~al.}(2008)\citenamefont
  {Srivastava}, \citenamefont {Htoon}, \citenamefont {Klimov},\ and\
  \citenamefont {Kono}}]{Srivastava2008}%
  \BibitemOpen
  \bibfield  {author} {\bibinfo {author} {\bibfnamefont {A.}~\bibnamefont
  {Srivastava}}, \bibinfo {author} {\bibfnamefont {H.}~\bibnamefont {Htoon}},
  \bibinfo {author} {\bibfnamefont {V.~I.}\ \bibnamefont {Klimov}}, \ and\
  \bibinfo {author} {\bibfnamefont {J.}~\bibnamefont {Kono}},\ }\href {\doibase
  10.1103/PhysRevLett.101.087402} {\bibfield  {journal} {\bibinfo  {journal}
  {Phys. Rev. Lett.}\ }\textbf {\bibinfo {volume} {101}},\ \bibinfo {pages}
  {087402} (\bibinfo {year} {2008})}\BibitemShut {NoStop}%
\bibitem [{\citenamefont {Slobodeniuk}\ and\ \citenamefont
  {Basko}(2016)}]{Slobo2016}%
  \BibitemOpen
  \bibfield  {author} {\bibinfo {author} {\bibfnamefont {A.~O.}\ \bibnamefont
  {Slobodeniuk}}\ and\ \bibinfo {author} {\bibfnamefont {D.~M.}\ \bibnamefont
  {Basko}},\ }\href {http://stacks.iop.org/2053-1583/3/i=3/a=035009} {\bibfield
   {journal} {\bibinfo  {journal} {2D Materials}\ }\textbf {\bibinfo {volume}
  {3}},\ \bibinfo {pages} {035009} (\bibinfo {year} {2016})}\BibitemShut
  {NoStop}%
\bibitem [{\citenamefont {Dery}\ and\ \citenamefont {Song}(2015)}]{dery2015}%
  \BibitemOpen
  \bibfield  {author} {\bibinfo {author} {\bibfnamefont {H.}~\bibnamefont
  {Dery}}\ and\ \bibinfo {author} {\bibfnamefont {Y.}~\bibnamefont {Song}},\
  }\href {\doibase 10.1103/PhysRevB.92.125431} {\bibfield  {journal} {\bibinfo
  {journal} {Phys. Rev. B}\ }\textbf {\bibinfo {volume} {92}},\ \bibinfo
  {pages} {125431} (\bibinfo {year} {2015})}\BibitemShut {NoStop}%
\bibitem [{\citenamefont {Chen}\ \emph {et~al.}(1988)\citenamefont {Chen},
  \citenamefont {Gil}, \citenamefont {Lefebvre},\ and\ \citenamefont
  {Mathieu}}]{chen1988}%
  \BibitemOpen
  \bibfield  {author} {\bibinfo {author} {\bibfnamefont {Y.}~\bibnamefont
  {Chen}}, \bibinfo {author} {\bibfnamefont {B.}~\bibnamefont {Gil}}, \bibinfo
  {author} {\bibfnamefont {P.}~\bibnamefont {Lefebvre}}, \ and\ \bibinfo
  {author} {\bibfnamefont {H.}~\bibnamefont {Mathieu}},\ }\href {\doibase
  10.1103/PhysRevB.37.6429} {\bibfield  {journal} {\bibinfo  {journal} {Phys.
  Rev. B}\ }\textbf {\bibinfo {volume} {37}},\ \bibinfo {pages} {6429}
  (\bibinfo {year} {1988})}\BibitemShut {NoStop}%
\bibitem [{\citenamefont {Andreani}\ and\ \citenamefont
  {Bassani}(1990)}]{andreani1990}%
  \BibitemOpen
  \bibfield  {author} {\bibinfo {author} {\bibfnamefont {L.~C.}\ \bibnamefont
  {Andreani}}\ and\ \bibinfo {author} {\bibfnamefont {F.}~\bibnamefont
  {Bassani}},\ }\href {\doibase 10.1103/PhysRevB.41.7536} {\bibfield  {journal}
  {\bibinfo  {journal} {Phys. Rev. B}\ }\textbf {\bibinfo {volume} {41}},\
  \bibinfo {pages} {7536} (\bibinfo {year} {1990})}\BibitemShut {NoStop}%
\bibitem [{\citenamefont {Castellanos-Gomez}\ \emph {et~al.}(2014)\citenamefont
  {Castellanos-Gomez}, \citenamefont {Buscema}, \citenamefont {Molenaar},
  \citenamefont {Singh}, \citenamefont {Janssen}, \citenamefont {van~der
  Zant},\ and\ \citenamefont {Steele}}]{gomez}%
  \BibitemOpen
  \bibfield  {author} {\bibinfo {author} {\bibfnamefont {A.}~\bibnamefont
  {Castellanos-Gomez}}, \bibinfo {author} {\bibfnamefont {M.}~\bibnamefont
  {Buscema}}, \bibinfo {author} {\bibfnamefont {R.}~\bibnamefont {Molenaar}},
  \bibinfo {author} {\bibfnamefont {V.}~\bibnamefont {Singh}}, \bibinfo
  {author} {\bibfnamefont {L.}~\bibnamefont {Janssen}}, \bibinfo {author}
  {\bibfnamefont {H.~S.~J.}\ \bibnamefont {van~der Zant}}, \ and\ \bibinfo
  {author} {\bibfnamefont {G.~A.}\ \bibnamefont {Steele}},\ }\href
  {http://stacks.iop.org/2053-1583/1/i=1/a=011002} {\bibfield  {journal}
  {\bibinfo  {journal} {2D Materials}\ }\textbf {\bibinfo {volume} {1}},\
  \bibinfo {pages} {011002} (\bibinfo {year} {2014})}\BibitemShut {NoStop}%
\bibitem [{\citenamefont {Mitioglu}\ \emph {et~al.}(2015)\citenamefont
  {Mitioglu}, \citenamefont {Plochocka}, \citenamefont {Granados~del Aguila},
  \citenamefont {Christianen}, \citenamefont {Deligeorgis}, \citenamefont
  {Anghel}, \citenamefont {Kulyuk},\ and\ \citenamefont {Maude}}]{mitoglunano}%
  \BibitemOpen
  \bibfield  {author} {\bibinfo {author} {\bibfnamefont {A.~A.}\ \bibnamefont
  {Mitioglu}}, \bibinfo {author} {\bibfnamefont {P.}~\bibnamefont {Plochocka}},
  \bibinfo {author} {\bibfnamefont {A.}~\bibnamefont {Granados~del Aguila}},
  \bibinfo {author} {\bibfnamefont {P.~C.~M.}\ \bibnamefont {Christianen}},
  \bibinfo {author} {\bibfnamefont {G.}~\bibnamefont {Deligeorgis}}, \bibinfo
  {author} {\bibfnamefont {S.}~\bibnamefont {Anghel}}, \bibinfo {author}
  {\bibfnamefont {L.}~\bibnamefont {Kulyuk}}, \ and\ \bibinfo {author}
  {\bibfnamefont {D.~K.}\ \bibnamefont {Maude}},\ }\href {\doibase
  10.1021/acs.nanolett.5b00626} {\bibfield  {journal} {\bibinfo  {journal}
  {Nano Lett.}\ }\textbf {\bibinfo {volume} {15}},\ \bibinfo {pages} {4387}
  (\bibinfo {year} {2015})}\BibitemShut {NoStop}%
\bibitem [{\citenamefont {Plechinger}\ \emph {et~al.}(2015)\citenamefont
  {Plechinger}, \citenamefont {Nagler}, \citenamefont {Kraus}, \citenamefont
  {Paradiso}, \citenamefont {Strunk}, \citenamefont {Sch\"uller},\ and\
  \citenamefont {Korn}}]{plechinger}%
  \BibitemOpen
  \bibfield  {author} {\bibinfo {author} {\bibfnamefont {G.}~\bibnamefont
  {Plechinger}}, \bibinfo {author} {\bibfnamefont {P.}~\bibnamefont {Nagler}},
  \bibinfo {author} {\bibfnamefont {J.}~\bibnamefont {Kraus}}, \bibinfo
  {author} {\bibfnamefont {N.}~\bibnamefont {Paradiso}}, \bibinfo {author}
  {\bibfnamefont {C.}~\bibnamefont {Strunk}}, \bibinfo {author} {\bibfnamefont
  {C.}~\bibnamefont {Sch\"uller}}, \ and\ \bibinfo {author} {\bibfnamefont
  {T.}~\bibnamefont {Korn}},\ }\href {\doibase 10.1002/pssr.201510224}
  {\bibfield  {journal} {\bibinfo  {journal} {Phys. Stat. Sol. RRL}\ }\textbf
  {\bibinfo {volume} {9}},\ \bibinfo {pages} {457} (\bibinfo {year}
  {2015})}\BibitemShut {NoStop}%
\bibitem [{\citenamefont {Shang}\ \emph {et~al.}(2015)\citenamefont {Shang},
  \citenamefont {Shen}, \citenamefont {Cong}, \citenamefont {Peimyoo},
  \citenamefont {Cao}, \citenamefont {Eginligil},\ and\ \citenamefont
  {Yu}}]{shang}%
  \BibitemOpen
  \bibfield  {author} {\bibinfo {author} {\bibfnamefont {J.}~\bibnamefont
  {Shang}}, \bibinfo {author} {\bibfnamefont {X.}~\bibnamefont {Shen}},
  \bibinfo {author} {\bibfnamefont {C.}~\bibnamefont {Cong}}, \bibinfo {author}
  {\bibfnamefont {N.}~\bibnamefont {Peimyoo}}, \bibinfo {author} {\bibfnamefont
  {B.}~\bibnamefont {Cao}}, \bibinfo {author} {\bibfnamefont {M.}~\bibnamefont
  {Eginligil}}, \ and\ \bibinfo {author} {\bibfnamefont {T.}~\bibnamefont
  {Yu}},\ }\href {\doibase 10.1021/nn5059908} {\bibfield  {journal} {\bibinfo
  {journal} {ACS Nano}\ }\textbf {\bibinfo {volume} {9}},\ \bibinfo {pages}
  {647} (\bibinfo {year} {2015})}\BibitemShut {NoStop}%
\bibitem [{\citenamefont {Li}\ \emph {et~al.}(2014{\natexlab{b}})\citenamefont
  {Li}, \citenamefont {Ludwig}, \citenamefont {Low}, \citenamefont {Chernikov},
  \citenamefont {Cui}, \citenamefont {Arefe}, \citenamefont {Kim},
  \citenamefont {van~der Zande}, \citenamefont {Rigosi}, \citenamefont {Hill},
  \citenamefont {Kim}, \citenamefont {Hone}, \citenamefont {Li}, \citenamefont
  {Smirnov},\ and\ \citenamefont {Heinz}}]{li2014}%
  \BibitemOpen
  \bibfield  {author} {\bibinfo {author} {\bibfnamefont {Y.}~\bibnamefont
  {Li}}, \bibinfo {author} {\bibfnamefont {J.}~\bibnamefont {Ludwig}}, \bibinfo
  {author} {\bibfnamefont {T.}~\bibnamefont {Low}}, \bibinfo {author}
  {\bibfnamefont {A.}~\bibnamefont {Chernikov}}, \bibinfo {author}
  {\bibfnamefont {X.}~\bibnamefont {Cui}}, \bibinfo {author} {\bibfnamefont
  {G.}~\bibnamefont {Arefe}}, \bibinfo {author} {\bibfnamefont {Y.~D.}\
  \bibnamefont {Kim}}, \bibinfo {author} {\bibfnamefont {A.~M.}\ \bibnamefont
  {van~der Zande}}, \bibinfo {author} {\bibfnamefont {A.}~\bibnamefont
  {Rigosi}}, \bibinfo {author} {\bibfnamefont {H.~M.}\ \bibnamefont {Hill}},
  \bibinfo {author} {\bibfnamefont {S.~H.}\ \bibnamefont {Kim}}, \bibinfo
  {author} {\bibfnamefont {J.}~\bibnamefont {Hone}}, \bibinfo {author}
  {\bibfnamefont {Z.}~\bibnamefont {Li}}, \bibinfo {author} {\bibfnamefont
  {D.}~\bibnamefont {Smirnov}}, \ and\ \bibinfo {author} {\bibfnamefont
  {T.~F.}\ \bibnamefont {Heinz}},\ }\href {\doibase
  10.1103/PhysRevLett.113.266804} {\bibfield  {journal} {\bibinfo  {journal}
  {Phys. Rev. Lett.}\ }\textbf {\bibinfo {volume} {113}},\ \bibinfo {pages}
  {266804} (\bibinfo {year} {2014}{\natexlab{b}})}\BibitemShut {NoStop}%
\bibitem [{\citenamefont {Arora}\ \emph
  {et~al.}(2015{\natexlab{b}})\citenamefont {Arora}, \citenamefont
  {Nogajewski}, \citenamefont {Molas}, \citenamefont {Koperski},\ and\
  \citenamefont {Potemski}}]{aroramose2}%
  \BibitemOpen
  \bibfield  {author} {\bibinfo {author} {\bibfnamefont {A.}~\bibnamefont
  {Arora}}, \bibinfo {author} {\bibfnamefont {K.}~\bibnamefont {Nogajewski}},
  \bibinfo {author} {\bibfnamefont {M.}~\bibnamefont {Molas}}, \bibinfo
  {author} {\bibfnamefont {M.}~\bibnamefont {Koperski}}, \ and\ \bibinfo
  {author} {\bibfnamefont {M.}~\bibnamefont {Potemski}},\ }\href {\doibase
  10.1039/C5NR06782K} {\bibfield  {journal} {\bibinfo  {journal} {Nanoscale}\
  }\textbf {\bibinfo {volume} {7}},\ \bibinfo {pages} {20769} (\bibinfo {year}
  {2015}{\natexlab{b}})}\BibitemShut {NoStop}%
\bibitem [{\citenamefont {Cadiz}\ \emph
  {et~al.}(2016{\natexlab{a}})\citenamefont {Cadiz}, \citenamefont {Robert},
  \citenamefont {Wang}, \citenamefont {Kong}, \citenamefont {Fan},
  \citenamefont {Blei}, \citenamefont {Lagarde}, \citenamefont {Gay},
  \citenamefont {Manca}, \citenamefont {Taniguchi}, \citenamefont {Watanabe},
  \citenamefont {Amand}, \citenamefont {Marie}, \citenamefont {Renucci},
  \citenamefont {Tongay},\ and\ \citenamefont {Urbaszek}}]{cadiz}%
  \BibitemOpen
  \bibfield  {author} {\bibinfo {author} {\bibfnamefont {F.}~\bibnamefont
  {Cadiz}}, \bibinfo {author} {\bibfnamefont {C.}~\bibnamefont {Robert}},
  \bibinfo {author} {\bibfnamefont {G.}~\bibnamefont {Wang}}, \bibinfo {author}
  {\bibfnamefont {W.}~\bibnamefont {Kong}}, \bibinfo {author} {\bibfnamefont
  {X.}~\bibnamefont {Fan}}, \bibinfo {author} {\bibfnamefont {M.}~\bibnamefont
  {Blei}}, \bibinfo {author} {\bibfnamefont {D.}~\bibnamefont {Lagarde}},
  \bibinfo {author} {\bibfnamefont {M.}~\bibnamefont {Gay}}, \bibinfo {author}
  {\bibfnamefont {M.}~\bibnamefont {Manca}}, \bibinfo {author} {\bibfnamefont
  {T.}~\bibnamefont {Taniguchi}}, \bibinfo {author} {\bibfnamefont
  {K.}~\bibnamefont {Watanabe}}, \bibinfo {author} {\bibfnamefont
  {T.}~\bibnamefont {Amand}}, \bibinfo {author} {\bibfnamefont
  {X.}~\bibnamefont {Marie}}, \bibinfo {author} {\bibfnamefont
  {P.}~\bibnamefont {Renucci}}, \bibinfo {author} {\bibfnamefont
  {S.}~\bibnamefont {Tongay}}, \ and\ \bibinfo {author} {\bibfnamefont
  {B.}~\bibnamefont {Urbaszek}},\ }\href {\doibase
  10.1088/2053-1583/3/4/045008} {\bibfield  {journal} {\bibinfo  {journal} {2D
  Materials}\ }\textbf {\bibinfo {volume} {3}},\ \bibinfo {pages} {045008}
  (\bibinfo {year} {2016}{\natexlab{a}})}\BibitemShut {NoStop}%
\bibitem [{\citenamefont {Cadiz}\ \emph
  {et~al.}(2016{\natexlab{b}})\citenamefont {Cadiz}, \citenamefont {Tricard},
  \citenamefont {Gay}, \citenamefont {Lagarde}, \citenamefont {Wang},
  \citenamefont {Robert}, \citenamefont {Renucci}, \citenamefont {Urbaszek},\
  and\ \citenamefont {Marie}}]{cadizacid}%
  \BibitemOpen
  \bibfield  {author} {\bibinfo {author} {\bibfnamefont {F.}~\bibnamefont
  {Cadiz}}, \bibinfo {author} {\bibfnamefont {S.}~\bibnamefont {Tricard}},
  \bibinfo {author} {\bibfnamefont {M.}~\bibnamefont {Gay}}, \bibinfo {author}
  {\bibfnamefont {D.}~\bibnamefont {Lagarde}}, \bibinfo {author} {\bibfnamefont
  {G.}~\bibnamefont {Wang}}, \bibinfo {author} {\bibfnamefont {C.}~\bibnamefont
  {Robert}}, \bibinfo {author} {\bibfnamefont {P.}~\bibnamefont {Renucci}},
  \bibinfo {author} {\bibfnamefont {B.}~\bibnamefont {Urbaszek}}, \ and\
  \bibinfo {author} {\bibfnamefont {X.}~\bibnamefont {Marie}},\ }\href
  {\doibase 10.1063/1.4954837} {\bibfield  {journal} {\bibinfo  {journal}
  {Appl. Phys. Lett.}\ }\textbf {\bibinfo {volume} {108}},\ \bibinfo {pages}
  {251106} (\bibinfo {year} {2016}{\natexlab{b}})}\BibitemShut {NoStop}%
\bibitem [{\citenamefont {Lezama}\ \emph {et~al.}(2015)\citenamefont {Lezama},
  \citenamefont {Arora}, \citenamefont {Ubaldini}, \citenamefont {Barreteau},
  \citenamefont {Giannini}, \citenamefont {Potemski},\ and\ \citenamefont
  {Morpurgo}}]{Ignacio2015}%
  \BibitemOpen
  \bibfield  {author} {\bibinfo {author} {\bibfnamefont {I.~G.}\ \bibnamefont
  {Lezama}}, \bibinfo {author} {\bibfnamefont {A.}~\bibnamefont {Arora}},
  \bibinfo {author} {\bibfnamefont {A.}~\bibnamefont {Ubaldini}}, \bibinfo
  {author} {\bibfnamefont {C.}~\bibnamefont {Barreteau}}, \bibinfo {author}
  {\bibfnamefont {E.}~\bibnamefont {Giannini}}, \bibinfo {author}
  {\bibfnamefont {M.}~\bibnamefont {Potemski}}, \ and\ \bibinfo {author}
  {\bibfnamefont {A.~F.}\ \bibnamefont {Morpurgo}},\ }\href {\doibase
  10.1021/nl5045007} {\bibfield  {journal} {\bibinfo  {journal} {Nano Lett.}\
  }\textbf {\bibinfo {volume} {15}},\ \bibinfo {pages} {2336} (\bibinfo {year}
  {2015})}\BibitemShut {NoStop}%
\end{thebibliography}%

\end{document}